\documentclass[a4paper,12pt]{article}

\usepackage[margin=2.5cm]{geometry}
\usepackage{graphicx}
\usepackage{amssymb}
\usepackage{multirow}
\usepackage{amsmath}
\usepackage{setspace}
\usepackage{caption}
\usepackage{subcaption}
\usepackage[colorlinks=true,bookmarks=true]{hyperref}
\usepackage{float}
\usepackage{mathrsfs}
\usepackage[nosort]{cite}

\hypersetup{citecolor=blue}
\newcommand{\dif}{\mathrm{d}}

\makeatletter
\renewcommand\section{\@startsection {section}{1}{\z@}%
	{-3.5ex \@plus -1ex \@minus -.2ex}%
	{2.3ex \@plus.2ex}%
	{\normalfont\bfseries}}
\renewcommand\subsection{\@startsection{subsection}{2}{\z@}%
	{-3.25ex\@plus -1ex \@minus -.2ex}%
	{1.5ex \@plus .2ex}%
	{\normalfont\it}}
\renewcommand\subsubsection{\@startsection{subsubsection}{3}{\z@}%
	{-3.25ex\@plus -1ex \@minus -.2ex}%
	{1.5ex \@plus .2ex}%
	{\normalfont}}
\makeatother

\begin{document}

\thispagestyle{empty}
\begin{flushright}
	
\end{flushright}
\vbox{}
\vspace{2cm}

\begin{center}
  {\LARGE{Rotating and accelerating black holes\\[2mm]
  with cosmological constant
  }}\\[16mm]
  {{Yu Chen,~~Cheryl Ng~~and~~Edward Teo}}
  \\[6mm]
    {\it Department of Physics,
      National University of Singapore, 
      Singapore 119260}\\[15mm]
\end{center}
\vspace{2cm}
	
\centerline{\bf Abstract}
\bigskip
\noindent We propose a new form of the rotating C-metric with cosmological constant, which generalises the form found by Hong and Teo for the Ricci-flat case. This solution describes the entire class of spherical black holes undergoing rotation and acceleration in dS or AdS space-time. The new form allows us to identify the complete ranges of coordinates and parameters of this solution. We perform a systematic study of its geometrical and physical properties, and of the various limiting cases that arise from it.

\newpage

\tableofcontents

\section{Introduction}

The Pleba\'nski--Demia\'nski solution \cite{Plebanski:1976gy} is a very general solution of the Einstein--Maxwell equations, that contains all the known single-black-hole solutions of this theory. Apart from the famous Kerr--Newman solution which describes a rotating and charged black hole, it contains extensions of the latter to include a cosmological constant, acceleration and NUT charge (see, e.g., \cite{Griffiths:2009dfa}). It is also known to contain large classes of black-hole solutions with non-spherical horizon topologies (see, e.g., \cite{Klemm:1997ea,Klemm:2014rda,Chen:2015zoa,Chen:2016rjt}).

Despite its generality, the Pleba\'nski--Demia\'nski solution can be written in a remarkably compact form. This form involves two related quartic functions, each of a single coordinate. The parameters of this solution are encoded in the coefficients of these two quartic functions. It has been traditional to take the parameters to be {\it the\/} coefficients themselves. However, starting from the work of \cite{Hong:2003gx,Hong:2004dm}, it was realised that a better choice would be to take the parameters to be the roots of the quartic functions. This leads to drastic simplifications when analysing the properties of the solution.

To illustrate this point, let us focus on the static limit of the Pleba\'nski--Demia\'nski solution. This special case is known as the C-metric, and describes a black hole undergoing an acceleration, but with no rotation or NUT charge. The C-metric is traditionally written in the form proposed by Kinnersley and Walker \cite{Kinnersley:1970zw}, who used a coordinate shift and rescaling to write it in the form (for simplicity, we only consider the uncharged case):
\begin{align}
\dif s^2&=\frac{1}{A^2(x-y)^2}\left[F(y)\dif t^2-\frac{\dif y^2}
{F(y)}+\frac{\dif x^2}{G(x)}+G(x)\dif\phi^2\right],\nonumber\\
  G(x)&=1-{x}^2-2 m A{x}^3,\nonumber\\
  F(y)&=\left(1-\frac{1}{\ell^2A^2}\right)-{y}^2-2 m A{y}^3.
\end{align}
Here $\ell$ is related to the cosmological constant $\Lambda$ by $\Lambda\equiv-3/\ell^2$, while $m$ and $A$ are parameters related to the mass and acceleration of the black hole, respectively. Since $G$ and $F$ are cubic functions, their roots are cumbersome to write down in terms of $\ell$, $m$ and $A$. Yet, it is important to know the form of these roots, since they encode the locations of the axes and Killing horizons of the space-time.

When the cosmological constant is zero, note that $G$ and $F$ share the same coefficients. In the case, Hong and Teo \cite{Hong:2003gx} realised that instead of using the above-mentioned coordinate freedom to set the linear coefficient of $G$ to zero, it is better to set it to a value such that $G$ can be written in the form:
\begin{align}
G(x)=(1-x^2)(1+2mAx)\,,
\end{align}
and similarly for $F$. In this form, their roots are trivial to read off: the two axes of the space-time are located at $x=\pm1$, while the acceleration and black-hole horizons are located at $y=-1$, $-\frac{1}{2mA}$, respectively. It is important to emphasize that this new form of the C-metric is physically equivalent to the traditional form of it. The difference lies only in how the solution is parameterised: the traditional form effectively uses the coefficients of the cubic functions as parameters, while the new form effectively uses their roots as parameters. 

When the cosmological constant is non-zero, the constant coefficients of $G$ and $F$ are different. This means that there is no longer a simple relationship between the roots of $G$ and those of $F$, and it is not immediately obvious how the form of \cite{Hong:2003gx} can be extended to this case. In fact, it was not until recently that the authors of \cite{Chen:2015vma} succeeded in doing so. Starting from the static limit of the Pleba\'nski--Demia\'nski solution, and assuming that the two cubic functions have two real roots each, they obtained a metric of the form:
\begin{align}
 \dif s^2&=\frac{\ell^2(a^2-1)(1-b^2)}{(x-y)^2}\left(Q(y)\dif t^2-\frac{\dif y^2}{Q(y)}+\frac{\dif x^2}{P(x)}+P(x)\dif\phi^2\right),\nonumber\\
 P(x)&=(x^2-1)[(a+b)(x-a-b)+ab+1]\,,\nonumber\\
 Q(y)&=(y-a)(y-b)[(a+b)y+ab+1]\,.\label{newform}
\end{align}
Note that the cosmological-constant parameter $\ell^2$ now appears as a conformal factor of the metric. The other two parameters of the solution, $a$ and $b$, are simply chosen to be the roots of the cubic function $Q$. A coordinate freedom can be used to set the two roots of the cubic function $P$ to be $\pm1$. The third root of $P$ and of $Q$ are then determined in terms of $a$ and $b$.\footnote{Recall that cubic polynomials either have exactly one or three real roots. Thus, the third root of $P$ and of $Q$ in this case are necessarily real ($a$ and $b$ are, of course, assumed to be real).} It can be checked that in the limit of zero cosmological constant ($\ell^2\rightarrow\pm\infty$ and $b\rightarrow-1$), the form of the C-metric in \cite{Hong:2003gx} is obtained.

Now, the two roots $x=\pm1$ represent axes of the space-time, while the two roots $y=a,b$ represent a black-hole and an acceleration horizon of the space-time. Thus the metric (\ref{newform}) describes a black hole undergoing an acceleration in de Sitter space-time \cite{Podolsky:2000pp,Dias:2003xp,Krtous:2003tc} or anti-de Sitter space-time \cite{Dias:2002mi,Podolsky:2003gm,Krtous:2005ej}. However, it is known that in AdS space-time, there exists a class of ``slowly accelerating'' black holes \cite{Podolsky:2002nk,Dias:2002mi,Krtous:2005ej}, whose space-times do {\it not\/} contain acceleration horizons. This class of black holes is not described by (\ref{newform}), for the simple reason that the form of $Q$ in (\ref{newform}) assumes the existence of an acceleration horizon (in addition to a black-hole horizon). 

Thus, to describe the class of slowly accelerating AdS black holes, one has to use a different form of (\ref{newform}) which does not assume that $Q$ has three real roots. Indeed, a form of the AdS C-metric in which $P$ and $Q$ are assumed to have one real root each was presented in \cite{Chen:2015zoa}. The other two roots of $P$ and of $Q$ may either be real or complex. The case where $P$ has three real roots and $Q$ has only one real root describes the slowly accelerating AdS black holes. We remark that the case where $P$ and $Q$ have only one real root each was also studied in \cite{Chen:2015zoa}, and was shown to describe slowly accelerating AdS black holes with a non-spherical horizon topology (this case was in fact the main focus of that paper).

It is instructive to recall the form of the AdS C-metric used in \cite{Chen:2015zoa}:
\begin{align}
 \dif s^2&=\frac{\ell^2}{(x-y)^2}\left(Q(y)\dif t^2-\frac{\dif y^2}{Q(y)}+\frac{\dif x^2}{P(x)}+P(x)\dif\phi^2\right),\cr
 P(x)&=(1+x)(1+\nu x+\mu x^2)\,,\cr
 Q(y)&=y\left[1+\nu+(\mu+\nu)y+\mu y^2\right].\label{newform2}
\end{align}
As can be seen, $P$ and $Q$ have one explicit real root each. A coordinate freedom can be used to set them to be $x=-1$ and $y=0$, respectively. The solution is parameterised by $\mu$ and $\nu$ (in addition to $\ell$), which appear as coefficients of the quadratic factor in $P$. Depending on the ranges of $\mu$ and $\nu$, the roots of this quadratic factor (as well as that in $Q$) may be complex. This is the reason why the roots were not used as parameters here.

The above two metrics (\ref{newform}) and (\ref{newform2}) show that there exist different forms of the C-metric with cosmological constant, that are adapted to different physical situations. The number of axes and Killing horizons in the space-time translate to the number of real roots of $P$ and $Q$. This would in turn determine how best to parameterise the solution. In general, it is possible to use a coordinate freedom to set two of the roots to specific values. The remaining real roots, if they exist, can then be used to parameterise the solution. Otherwise, suitable polynomial coefficients would have to be used as parameters.

In this paper, we will be primarily interested in {\it spherical\/} black holes undergoing an acceleration in dS or AdS space-time. The condition that these black holes are spherical immediately implies that $P$ must have two (and hence three) real roots. On the other hand, we should not impose the same on $Q$, since we have mentioned that there exists a class of slowly accelerating black holes in AdS space-time which do not have acceleration horizons. Thus, we impose that $Q$ has (at least) one real root. 

The two real roots of $P$ can, as usual, be taken to be $\pm1$. The only other real roots that we know exist at this stage are the third root of $P$ and the one real root of $Q$. It might seem natural to parameterise the solution in terms of these two roots. But unlike the above two cases, these two roots belong to different functions and the resulting parameterisations are not very elegant. After some trial and error, the simplest parameterisation of this solution turns out to be given by (\ref{metric_static}) below. As can be seen from the first equation of (\ref{PQ_static}), the parameter $c$ is directly related to the third root of $P$. However, the parameter $b$ has no direct relation to the real root of $Q$; instead, it has been chosen to be simply the constant coefficient of $Q$. In the light of what was discussed above, this might seem regressive since there are no simple expressions for the roots of $Q$. In choosing this parameterisation, we were also driven by a consideration hitherto not mentioned: that the chosen parameterisation should admit a simple generalisation to the {\it rotating\/} case. It turns out that the metric (\ref{metric_static}) does have an elegant rotating generalisation.

It is our aim in this paper to show that the metric (\ref{metric_static}), and its rotating generalisation, provides a useful form to describe the entire class of spherical black holes undergoing an acceleration in dS or AdS space-time. To this end, we will show how it can be used to build up a complete picture of the parameter space of solutions. This is something that has proved very difficult to do in previous forms of this solution (e.g., \cite{Podolsky:2006px}). With the parameter space identified, we are then able to describe the various parts and boundaries of it. In particular, we mention that one boundary of the parameter space corresponds to the ``black bottle'' solution that was studied in \cite{Chen:2016rjt}.

This paper is organised as follows: We begin in Sec.~\ref{sec_static} with a study of the static metric (\ref{metric_static}). We find the appropriate coordinate and parameter ranges, and study its geometrical and physical properties. We also discuss various special cases of this solution. In Sec.~\ref{sec_Crotating}, we turn our attention to the rotating generalisation of (\ref{metric_static}), and build upon the analysis of the static case. The paper concludes with a discussion of a few possible avenues for future work.

\section{Static accelerating black hole}
\label{sec_static}

The metric describing the static accelerating spherical black hole can be written as
\begin{align}
\dif s^2&=\frac{\ell^2(1-b)}{(x-y)^2}\bigg[Q(y)\dif t^2-\frac{\dif y^2}{Q(y)}+\frac{\dif x^2}{P(x)}+P(x)\dif \phi^2\bigg]\,,\nonumber\\
	P(x)&=1+cx-x^2-cx^3,\qquad
	Q(y)=b+cy-y^2-cy^3.
\label{metric_static}
\end{align}
We note that the functions $P$ and $Q$ can also be written as
\begin{align}
P(x)=(1-x^2)(1+cx)\,,\qquad Q(y)=P(y)-(1-b)\,.
\label{PQ_static}
\end{align}
This solution has three parameters: $\ell$, $b$ and $c$. The first parameter is related to the cosmological constant $\Lambda$ by
\begin{align}
\Lambda=-\frac{3}{\ell^2}\,.
\end{align}
It can be seen that the de Sitter (dS) case $\Lambda>0$ corresponds to $\ell^2<0$, while the anti-de Sitter (AdS) case $\Lambda<0$ corresponds to $\ell^2>0$. The Ricci-flat case $\Lambda=0$ is recovered in the limit $\ell^2\rightarrow\pm\infty$.

\subsection{Coordinate and parameter ranges}

Of the three roots of the function $P$, two can be fixed at $x=\pm1$ by using the rescaling and translational symmetries of the metric (\ref{metric_static}); see, e.g., \cite{Chen:2015vma}. They represent the two axes of the space-time, and we will only be interested in the region of the space-time lying between them:
\begin{align}
-1<x<+1\,.
\label{range_x}
\end{align}
Since $P(x=0)=1>0$, $P$ should remain positive within the above range. This means that the parameter $c$ must lie in the range
\begin{align}
-1\le c\le 1\,.
\label{range_c0}
\end{align}

In order to have the correct Lorentzian signature ($-$+++) for the metric (\ref{metric_static}), we require $Q<0$. Furthermore, the overall constant factor $\ell^2(1-b)$ of the metric must be positive. So we necessarily have either
\begin{align}
\ell^2>0\,,\qquad b<1\,,
\label{b_AdS}
\end{align}
or
\begin{align}
\ell^2<0\,,\qquad b>1\,.
\label{b_dS}
\end{align}
The former corresponds to the AdS case, while the latter corresponds to the dS case. The Ricci-flat limit can be obtained from either case, by letting
\begin{align}
\ell^2\rightarrow\pm\infty\,,\qquad b\rightarrow 1\,.
\label{b_Ricci-flat}
\end{align}
Thus, we see that the value of the parameter $b$ with respect to 1 is tied directly to the sign of the cosmological constant. This is a nice property of the present parameterisation.

We now turn to the range of $y$. We note from (\ref{metric_static}) that conformal infinity of the space-time lies at $x=y$. Due to the reflection symmetry\footnote{\label{reflection}This refers to the property of the metric (\ref{metric_static}) that it remains invariant under the transformation $(x,y,t,\phi,c)\rightarrow-(x,y,t,\phi,c)$. A similar property holds for the metrics (\ref{metric_crotating}) and (\ref{metric_charged}) below.} of the metric (\ref{metric_static}), we can assume 
\begin{align}
y<x\,.
\end{align}
Thus, in an $x$-$y$ plot, the region of interest always lies {\it below\/} the curve $y=x$. 

We also require that the region of interest does not contain any curvature singularities. Now it can be checked that the metric (\ref{metric_static}) contains curvature singularities at $x,y=\pm\infty$. In particular, the curvature singularity at $y=-\infty$ can be avoided by demanding that the function $Q$ admits at least one real root $y_1$ satisfying $y_1<-1$, which represents the event horizon of a black hole. This event horizon has a spherical topology, and will enclose the curvature singularity $y=-\infty$. By restricting $y$ to the range
\begin{align}
y>y_1\,,
\end{align}
we will be focussing on the static region {\it outside\/} the black hole.

Recall that we should have $Q<0$ in this region. This means that $Q$ should satisfy
\begin{align}
Q(y\,{=}\,y_1)=0\,,\qquad Q'(y\,{=}\,y_1)<0\,,
\label{condition_existence_BH}
\end{align}
at $y=y_1$. These conditions in fact imply that the parameter $c$ is non-negative. For a negative $c$ satisfying (\ref{range_c0}), $P$ has a positive leading (cubic) coefficient and admits a third root lying above 1. Since the function $Q$ can be obtained from $P$ by a constant shift, we see that the conditions (\ref{condition_existence_BH}) cannot be simultaneously satisfied. Hence, the range of the parameter $c$ is given by
\begin{align}
0\leq c\leq 1\,.
\label{range_c_static}
\end{align}
With this range of $c$, the third root of $P$ lies at or below $-1$.

We now turn to the root structure of $Q$, which will then fix the range of the coordinate $y$ and hence the entire $(x,y)$ coordinate range of the region of interest. We shall refer to this coordinate range as the domain of the space-time. The domain of the space-time will encode useful information about its physical properties.

\subsubsection{Domain structure}
\label{static_domains}

\begin{figure}[t]
\begin{center}
 \begin{subfigure}[b]{0.4\textwidth}
  \centering
  \includegraphics[scale=0.45]{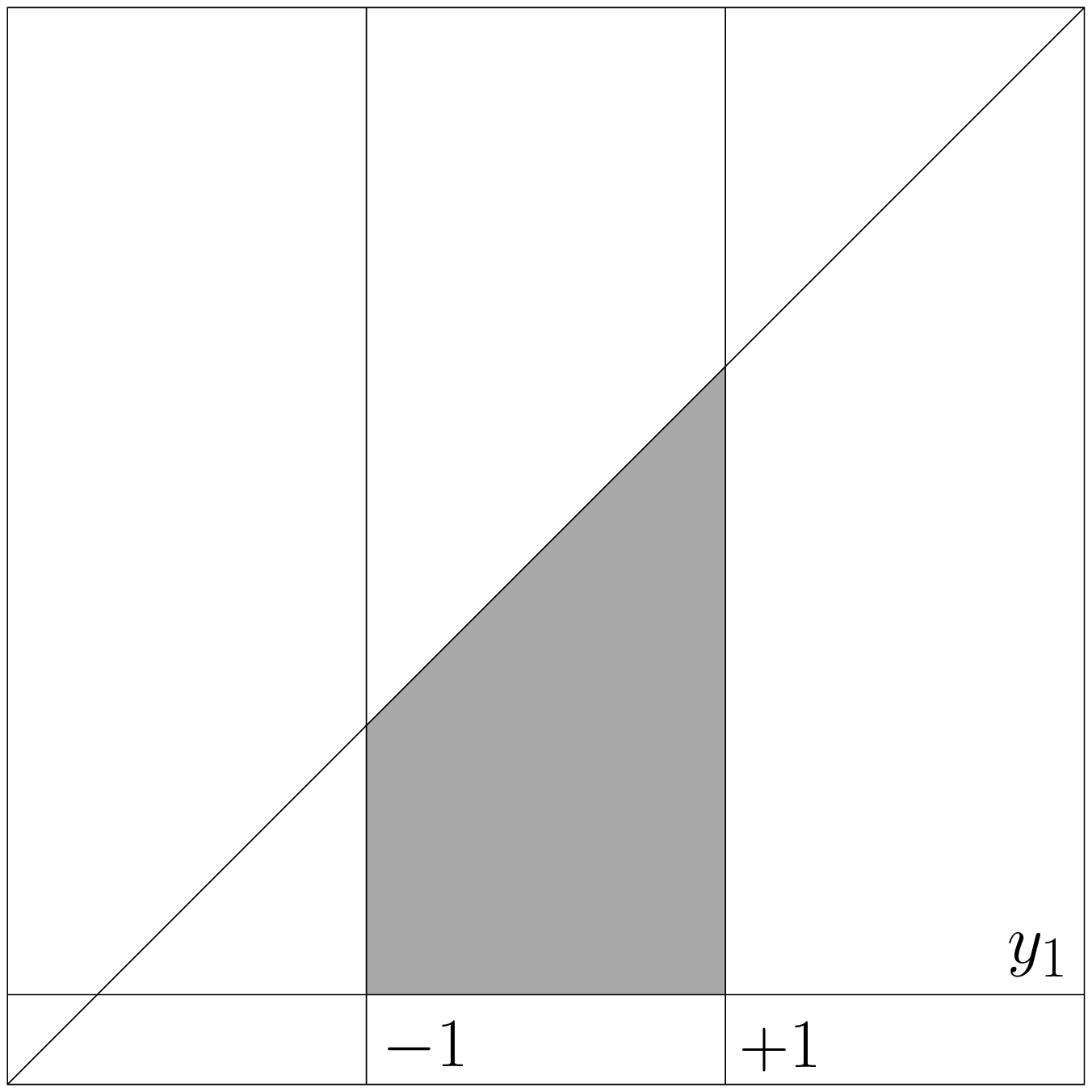}
  \caption{}
  \label{fig_domain_a}
 \end{subfigure}\vspace{6pt}
~~~ \begin{subfigure}[b]{0.4\textwidth}
  \centering
  \includegraphics[scale=0.45]{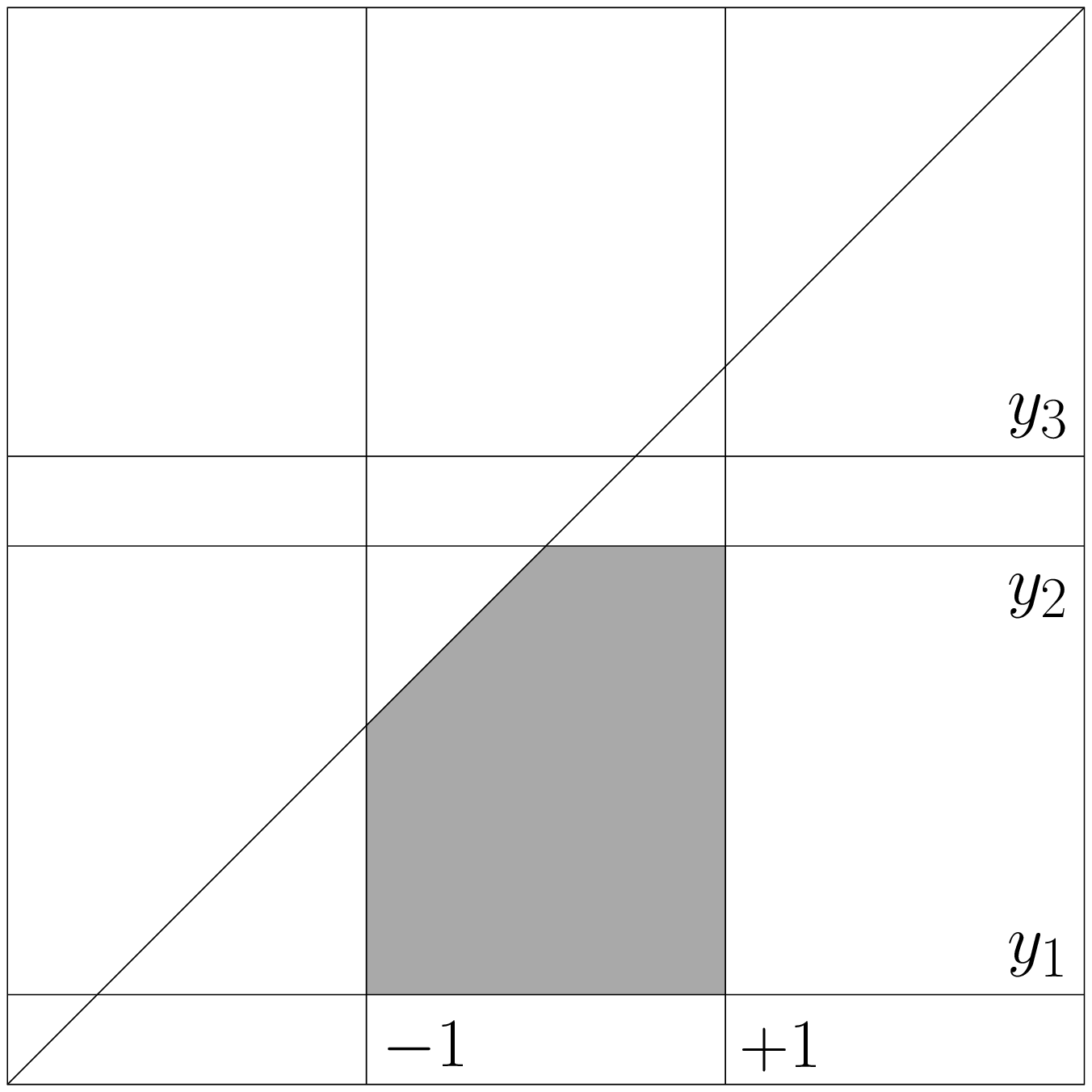}
  \caption{}
  \label{fig_domain_b}
 \end{subfigure}
 \begin{subfigure}[b]{0.4\textwidth}
  \centering
  \includegraphics[scale=0.45]{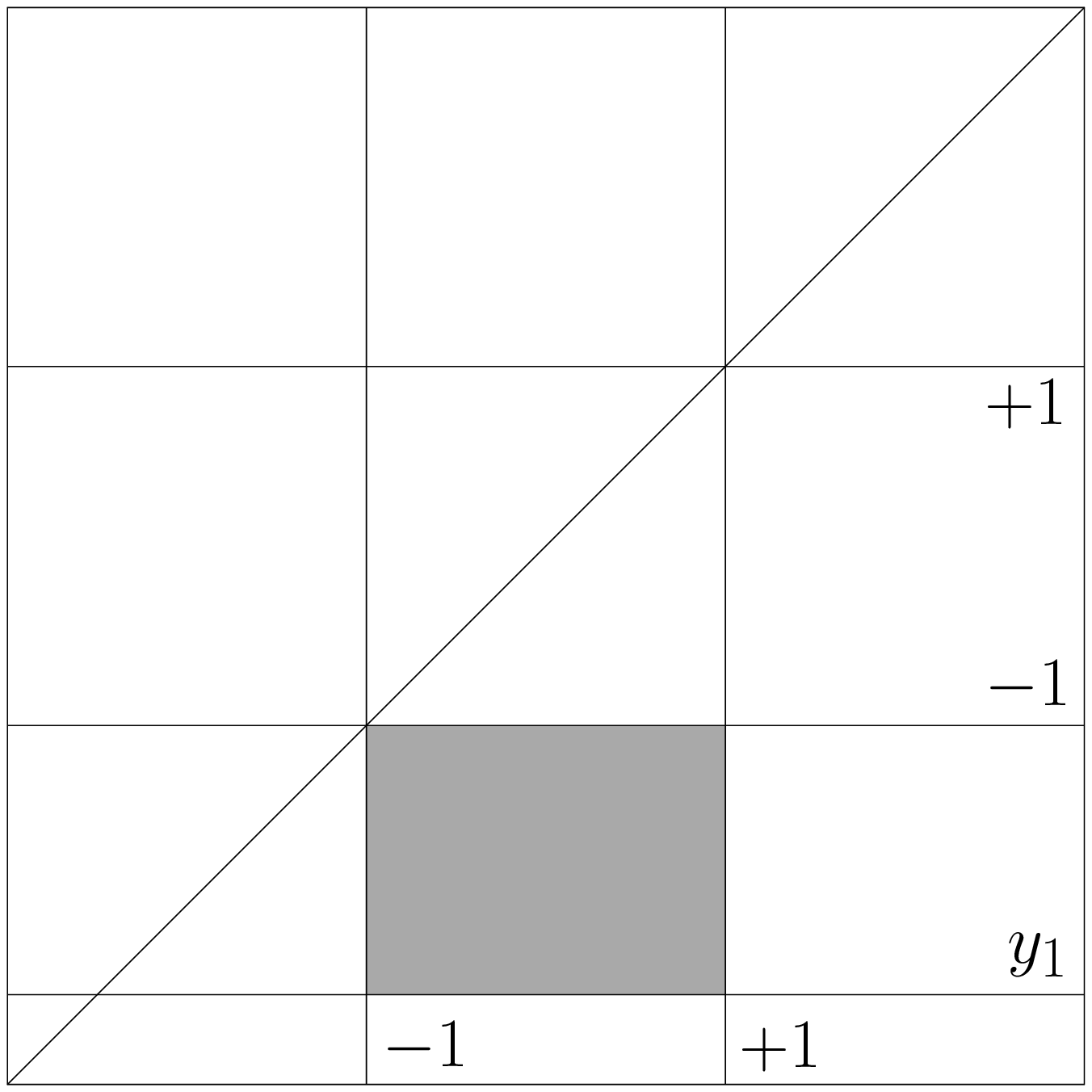}
  \caption{}
  \label{fig_domain_c}
 \end{subfigure}\hspace{12pt}
 \begin{subfigure}[b]{0.4\textwidth}
  \centering
  \includegraphics[scale=0.45]{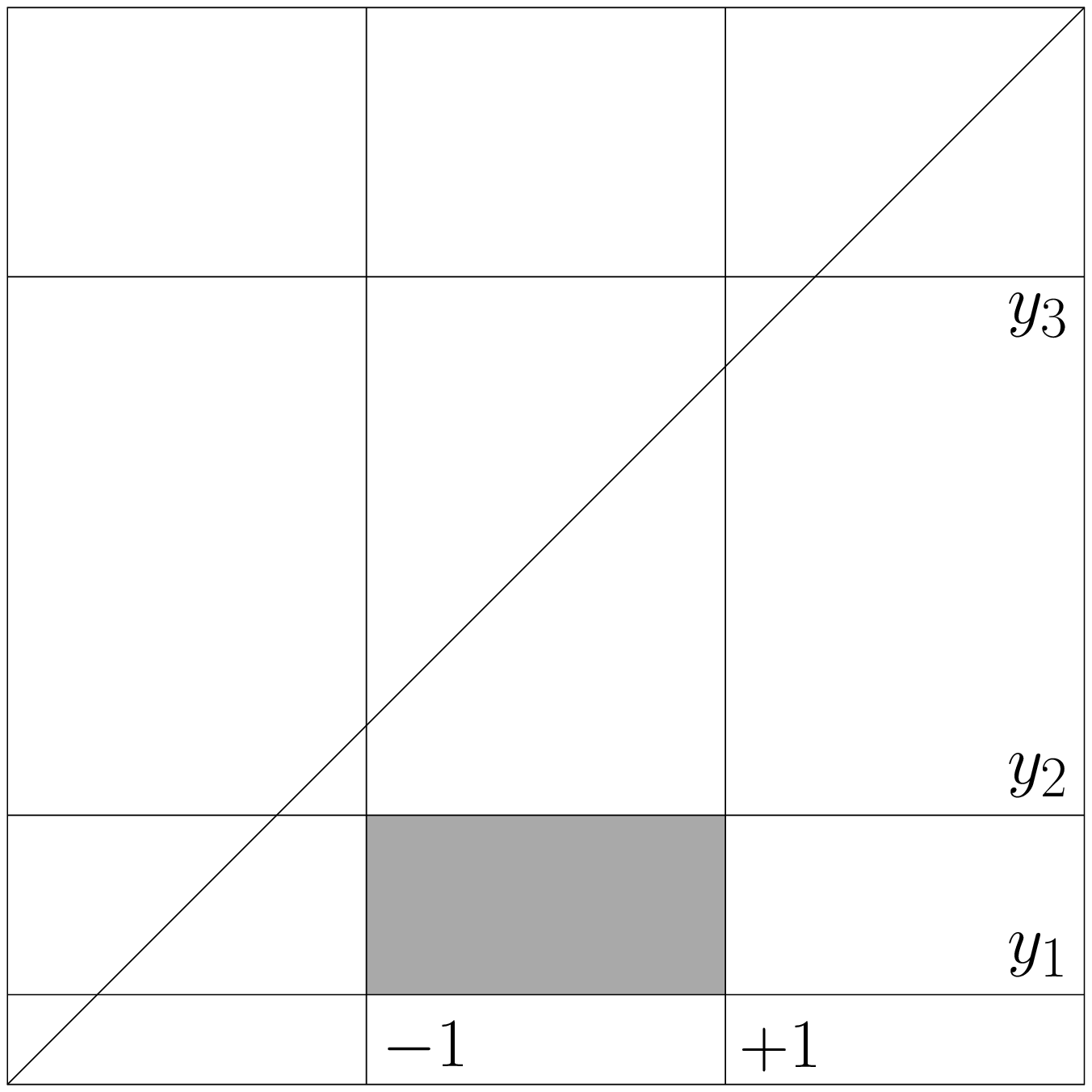}
  \caption{}
  \label{fig_domain_d}
 \end{subfigure}
\end{center}
\caption{The domains of the static accelerating black holes described by (\ref{metric_static}): In (a), $Q$ has only one real root, while in (b), (c) and (d), $Q$ admits three real roots. The first domain describes a black hole without acceleration horizon, which only occurs in the AdS case. The latter three domains describe black holes with acceleration horizons for the AdS, Ricci-flat and dS cases, respectively.}
\label{fig_domains}
\end{figure}

The root structure of $Q$ is largely determined by the value of the parameter $b$. For a fixed $c$ in the range (\ref{range_c_static}), and in view of (\ref{PQ_static}), we see that for a sufficiently negative $b$, $Q$ has only one root $y_1$. As mentioned above, this root should be identified as the location of the black-hole horizon. The coordinates in this case thus take the ranges
\begin{align}
-1<x<+1\,,\qquad y_1<y<x\,.
\label{coor_range1}
\end{align}
This domain can be visualised in an $x$-$y$ plot as shown in Fig.~\ref{fig_domain_a}. It describes the region outside a black hole in AdS space-time, which extends to conformal infinity.

Now if we increase $b$ above a certain critical value while still keeping it below 1, $Q$ will have three real roots $y_{1,2,3}$ satisfying
\begin{align}
-\infty<y_1<-1<y_2\leq y_3<+1\,.
\end{align}
The root $y_2$ now represents an acceleration horizon of the space-time. If we restrict ourselves to the static region of the space-time between the two Killing horizons, the coordinate ranges are
\begin{align}
-1<x<+1\,,\qquad y_1<y<x\,,\qquad y<y_2\,.
\label{coor_range2}
\end{align}
This domain is shown in Fig.~\ref{fig_domain_b}. It describes an accelerating black hole in AdS space-time. More specifically, it describes the static region of space-time between the black-hole and acceleration horizons.

When $b=1$ is reached, the roots of $Q$ will satisfy
\begin{align}
-\infty<y_1=-\frac{1}{c}<-1=y_2<+1=y_3\,.
\end{align}
The coordinate ranges are thus
\begin{align}
y_1<y<-1<x<+1\,.
\label{coor_range3}
\end{align}
This domain is shown in Fig.~\ref{fig_domain_c}. It describes a Ricci-flat accelerating black hole.

When $b$ is increased beyond 1 while still keeping it below a certain critical value, the roots of $Q$ will satisfy
\begin{align}
-\infty<y_1<y_2<-1<+1<y_3\,.
\end{align}
The coordinate ranges are thus
\begin{align}
-1<x<+1\,,\qquad y_1<y<y_2\,.
\label{coor_range4}
\end{align}
This domain is shown in Fig.~\ref{fig_domain_d}. It describes an accelerating black hole in dS space-time.

\subsubsection{Parameter space}

\begin{figure}
 	\begin{center}
 		\includegraphics[scale=1.2]{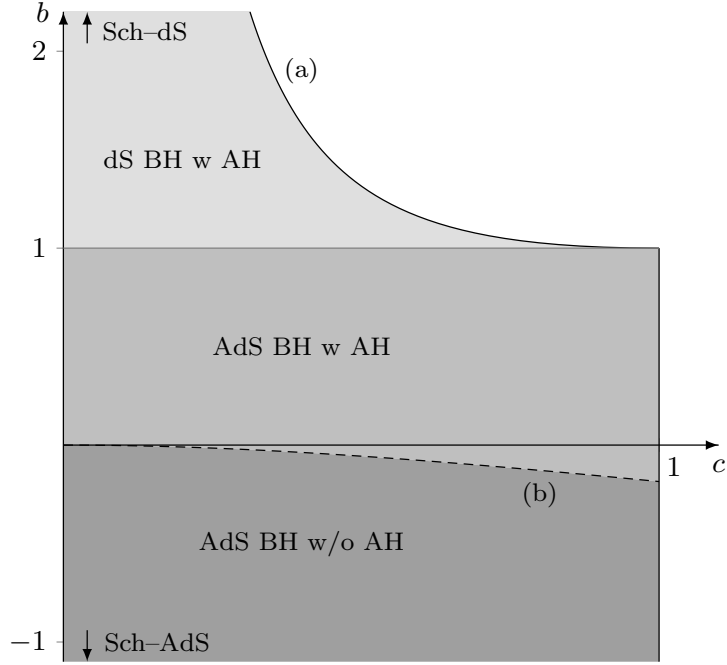}
 	\end{center}
 	\caption{The parameter space of the static accelerating black holes described by (\ref{metric_static}). It is bounded by three curves: $c=0$, $c=1$ and the curve labelled by (a). The $b=1$ line and the dashed curve (b) divide the parameter space into three distinct regions, corresponding to the domains depicted in Figs.~\ref{fig_domain_a}, \ref{fig_domain_b} and \ref{fig_domain_d}. Points on the $b=1$ line correspond to the Ricci-flat case in Fig.~\ref{fig_domain_c}.}
 	\label{fig_param_space1}
\end{figure}

We now describe the full parameter space of the metric (\ref{metric_static}). It turns out to be bounded by three curves: the lines $c=0$ and $c=1$, and the curve labelled by (a) as shown in Fig.~\ref{fig_param_space1}. Note that the $c=0$ line meets the other two curves at $b=\pm\infty$.
As we will discuss in more detail in Secs.~\ref{sec_static_massless} and \ref{sec_static_bottle} respectively, $c=0$ corresponds to the massless limit of the black hole, while $c=1$ corresponds to the limit in which the black hole becomes a black bottle. The curve (a) corresponds to the situation in which $y_1=y_2$ in Fig.~\ref{fig_domain_d}. The Lorentzian region of interest vanishes in this limit, and no static space-time exists beyond this curve. This special case will be discussed in Sec.~\ref{sec_static_extremal1}.

The dashed curve (b) in Fig.~\ref{fig_param_space1} corresponds to the situation in which $y_2=y_3$ in Fig.~\ref{fig_domain_b}. Physically, it corresponds to the acceleration horizon becoming extremal, and it separates the AdS black holes with acceleration horizons (Fig.~\ref{fig_domain_b}) from those without acceleration horizons (Fig.~\ref{fig_domain_a}). This limit will be discussed in Sec.~\ref{sec_static_extremal2}. On the other hand, the line $b=1$ separates the AdS black holes with acceleration horizons (Fig.~\ref{fig_domain_b}) from the dS black holes with acceleration horizons (Fig.~\ref{fig_domain_d}). The line itself corresponds to Ricci-flat black holes with acceleration horizons (Fig.~\ref{fig_domain_c}). This special case will be discussed in Sec.~\ref{sec_static_C}.

Finally, we remark that the Schwarzschild-dS and Schwarzschild-AdS black holes can be recovered in the scaling limit $b\rightarrow\pm\infty$ and $c\rightarrow0$. In Fig.~\ref{fig_param_space1}, this corresponds to approaching the top-left and bottom-left corners of the parameter space, where the $c=0$ line meets the other two boundary curves. These two special cases will be discussed in Sec.~\ref{sec_static_Sch}.

\subsection{Geometrical and physical properties}

Having obtained the possible domains of the metric (\ref{metric_static}) in Sec.~\ref{static_domains}, we now turn to a study of the boundary of each domain. This is most elegantly done using the rod-structure formalism; see, e.g., \cite{Harmark:2004rm,Hollands:2007aj,Chen:2010zu,Armas:2011ed} for more details of this formalism. In particular, we shall follow the formalism of \cite{Chen:2010zu} most closely here. A discussion of the horizon geometries based on the rod structure then follows.

\subsubsection{Rod structure}

We first consider the case in which an acceleration horizon is present in the space-time, corresponding to one of the domains in Fig.~\ref{fig_domain_b}, \ref{fig_domain_c} or \ref{fig_domain_d}. The rod structure in this case is
\begin{enumerate}
	\item Rod 1: a semi-infinite space-like rod located at $(x\,{=}\,{-1},y_1\,{\leq}\,y\,{<}\,{-1})$, with direction
	\begin{align}
          \label{k1}
		k_1=\frac{1}{\kappa_{\rm E1}}(0,1)\,,\qquad\kappa_{\rm E1}=1-c\,;
	\end{align}
	
	\item Rod 2: a finite time-like rod located at $(-1\,{\leq}\,x\,{\leq}\,{+1},y\,{=}\,y_1)$, with direction
	\begin{align}
          \label{k2}
		k_2=\frac{1}{\kappa_2}(1,0)\,,\qquad\kappa_2=-\frac{1}{2}\frac{\dif Q}{\dif y}\bigg|_{y=y_1}\,;
	\end{align}
	
	\item Rod 3: a finite space-like rod located at $(x\,{=}\,{+1},y_1\,{\leq}\,y\,{\leq}\,y_2)$, with direction
	\begin{align}
          \label{k3}
		k_3=\frac{1}{\kappa_{\rm E3}}(0,1)\,,\qquad\kappa_{\rm E3}=1+c\,;
	\end{align}
	
	\item Rod 4: a time-like rod located at $(\max(y_2,-1)\,{\leq}\,x\,{\leq}\,{+1},y\,{=}\,y_2)$, with direction
	\begin{align}
          \label{k4}
		k_4=\frac{1}{\kappa_4}(1,0)\,,\qquad \kappa_4=\frac{1}{2}\frac{\dif Q}{\dif y}\bigg|_{y=y_2}\,.
	\end{align}
\end{enumerate}
Since Rods 1 and 3 are space-like, they represent axes in the space-time. On the other hand, Rods 2 and 4 are time-like and represent Killing horizons in the space-time; they represent the black-hole and acceleration horizons respectively. Note that Rod 4 is semi-infinite in extent in the AdS and Ricci-flat cases, and is finite in extent in the dS case.

In the case when the acceleration horizon is absent, corresponding to the domain in Fig.~\ref{fig_domain_a}, Rod 4 has to be removed from the above rod structure. In this case, Rod 3 extends to conformal infinity, and has the coordinate range $(x\,{=}\,{+1},y_1\,{\leq}\,y\,{<}\,{+1})$. This, of course, only occurs in the AdS case.

From the rod direction (\ref{k1}), we see that the coordinate identification
\begin{align}
	(t,\phi)\rightarrow \Big(t,\phi+\frac{2\pi}{1-c}\Big)\,,
	\label{identification_static1}
\end{align}
has to be made to avoid a conical singularity along the axis $x=-1$. On the other hand, we see from (\ref{k3}) that the coordinate identification
\begin{align}
	(t,\phi)\rightarrow \Big(t,\phi+\frac{2\pi}{1+c}\Big)\,,
	\label{identification_static2}
\end{align}
has to be made to avoid a conical singularity along the axis $x=+1$. In view of the allowed range of $c$ given by (\ref{range_c_static}), these identifications cannot be made simultaneously unless $c=0$. If we make the identification (\ref{identification_static1}), the space-time contains a strut with a conical excess along the axis $x=+1$; if we make the identification (\ref{identification_static2}), the space-time contains a cosmic string with a conical deficit along the axis $x=-1$.

\subsubsection{Horizon geometries}

To study the geometry of the black-hole horizon represented by Rod 2, it is convenient to reparameterise the solution in terms of $c$ and $y_1$. This amounts to writing $b$ as
\begin{align}
b=cy_1(y_1^2-1)+y_1^2\,.
\end{align}
For a constant time slice, the induced metric on the horizon is
\begin{align}
\dif s_{\text{BH}}^2=\frac{\ell^2(1-y_1^2)(1+cy_1)}{(x-y_1)^2}\bigg[\frac{\dif x^2}{{(1-x^2)(1+cx)}}+{(1-x^2)(1+cx)}\dif \phi^2\bigg]\,.
\label{horizon_geometry_static}
\end{align}
We note that $\ell^2(1+cy_1)<0$, so this metric is positive semi-definite.

It can be checked that the induced metric is regular at the north or south pole $x=\pm 1$, if the corresponding identification (\ref{identification_static2}) or (\ref{identification_static1}) is made, respectively. For definiteness, we choose to make the identification (\ref{identification_static2}), so that the north pole of the horizon is regular. This leaves a conical singularity at the south pole, with a deficit angle given by
\begin{align}
\delta=\frac{4\pi c}{1+c}\,.
\label{conical_deficit_static}
\end{align}
With the identification (\ref{identification_static2}), the area of the horizon is
\begin{align}
A_{\text{BH}}&=-2\ell^2(1+cy_1)\Delta\phi\nonumber\\
&=-\frac{4\pi\ell^2(1+cy_1)}{1+c}\,.
\end{align}

An important geometrical quantity characterising the horizon is its scalar curvature, which can be calculated to be
\begin{align}
R(x)=-\frac{2}{\ell^2}+\frac{2c(x-y_1)^3}{\ell^2(1-y_1^2)(1+cy_1)}\,.
\end{align}
At the north and south poles, it is 
\begin{align}
R(x\,{=}\,{+1})&=-\frac{2}{\ell^2}\,\frac{1-c+(1+3c)y_1}{(1+y_1)(1+cy_1)}>0\,,\cr
R(x\,{=}\,{-1})&=-\frac{2}{\ell^2}\,\frac{2(1+cy_1)-(1-c)(1+y_1)}{(1-y_1)(1+cy_1)}\,.
\end{align}
It can be checked that the scalar curvature is always positive at the north pole. At the south pole however, it is positive in the dS and Ricci-flat cases, but can have either sign in the AdS case. Since $R$ is a monotonically increasing function of $x$, we conclude that it is positive everywhere on the horizon if it is positive at the south pole.

\begin{figure}[t]
\begin{center}
 \begin{subfigure}[b]{0.32\textwidth}
  \centering
  \includegraphics[height=2in,angle=-90]{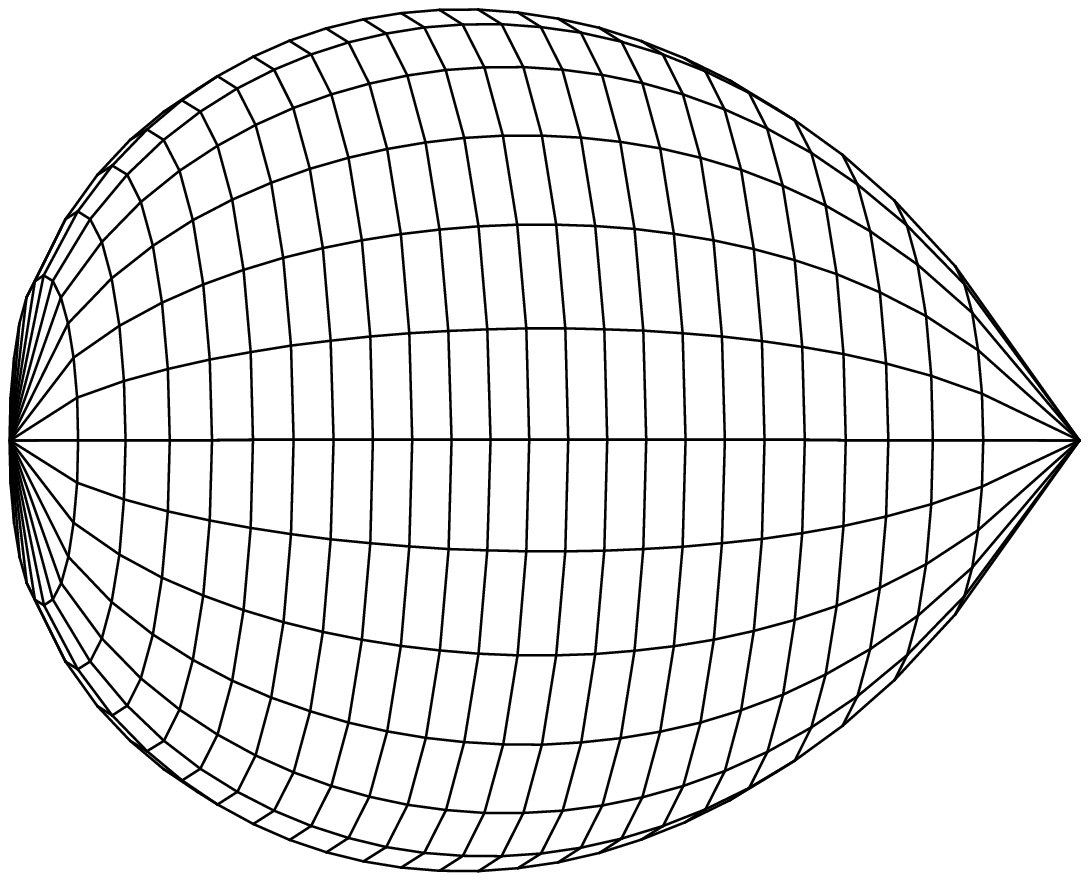}
  \caption{}
  \label{fig_horizon_a}
 \end{subfigure}\vspace{6pt}
 \begin{subfigure}[b]{0.32\textwidth}
  \centering
  \includegraphics[height=1.5in,angle=-90]{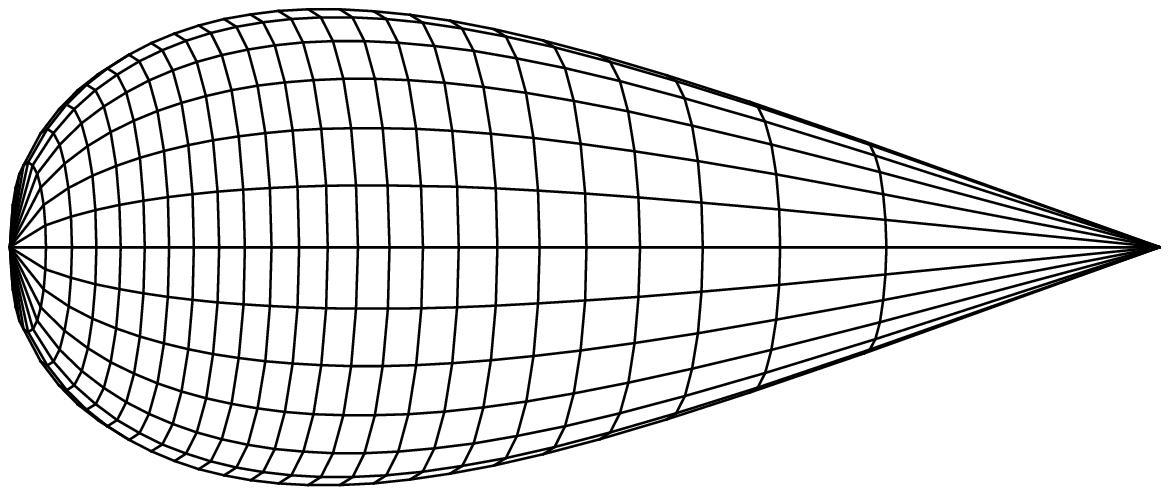}
  \caption{}
  \label{fig_horizon_b}
 \end{subfigure}
 \begin{subfigure}[b]{0.32\textwidth}
  \centering
  \includegraphics[height=1in,angle=-90]{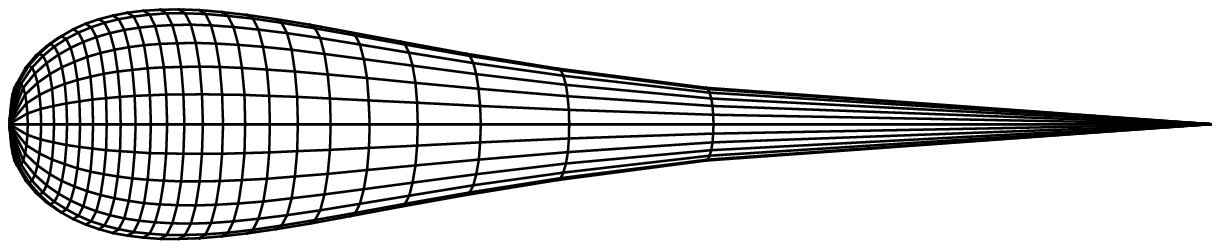}
  \caption{}
  \label{fig_horizon_c}
 \end{subfigure}
\end{center}
\caption{The horizon geometry (\ref{horizon_geometry_static}) of the static accelerating black hole, as embedded as a surface of revolution in a three-dimensional Euclidean space. In (a), (b) and (c), $R(x\,{=}\,{-1})$ is positive, zero and negative, respectively. The south pole in each case is attached to a cosmic string that extends to infinity at the bottom of the figure.}
\label{fig_horizon}
\end{figure}

It is instructive to embed the horizon as a surface of revolution in a three-dimensional Euclidean space, in order to visualise its geometry. We remark that this is not always possible if the scalar curvature is negative, although it turns out to be possible for the cases considered here. The embedding diagrams are shown in Fig.~\ref{fig_horizon}, for each possible sign for $R(x\,{=}\,{-1})$, as well as for the case in which it is zero. In all three cases, it is clear that the north pole is regular, just as it is clear that there is a conical singularity at the south pole. The difference between them lies in the concavity or convexity of the horizon geometry as the south pole is approached.

It is also clear from Fig.~\ref{fig_horizon} that the horizon becomes more elongated as $R(x\,{=}\,{-1})$ is decreased. In fact, in the limit $c\rightarrow1$ where $R(x\,{=}\,{-1})$ becomes the most negative, the south pole of the horizon extends to infinity, and the black hole becomes a black bottle \cite{Chen:2016rjt} (c.f.\ Sec.~\ref{sec_static_bottle}). At the other extreme, there is a limit in which $c\rightarrow0$ such that the horizon becomes perfectly spherical. This is the zero-acceleration limit (c.f.\ Sec.~\ref{sec_static_Sch}).

The geometry of the acceleration horizon can be analysed along similar lines, although we will not present the details here. We merely comment that in the AdS and Ricci-flat cases, the acceleration horizon extends to conformal infinity, and has the topology of a plane. However, in the dS case, it remains compact and has a spherical topology. It in fact surrounds and encloses the black-hole horizon, and is also known as a cosmological horizon.

When there are two Killing horizons present in the space-time, it is natural to ask if thermal equilibrium can be achieved between them. Recall that the temperature of a Killing horizon is given in terms of its surface gravity by $T=\frac{\kappa}{2\pi}$. From the expressions of the surface gravities of the black-hole and acceleration horizons in (\ref{k2}) and (\ref{k4}), it can be shown that $\kappa_2>\kappa_4$. Thus
\begin{align}
T_{\text{BH}}>T_{\text{AH}}\,,
\end{align}
i.e., the black-hole horizon is always hotter than the acceleration horizon.

\subsection{Special cases}
\label{sec_special_cases_static}

In this section, we shall study several important special cases of the metric (\ref{metric_static}). In the process, we will obtain a better understanding of the parameters of the solution, and the various boundaries and parts of the parameter space in Fig.~\ref{fig_param_space1}.

\subsubsection{Massless limit}
\label{sec_static_massless}

In the limit of vanishing $c$:
\begin{align}
c\rightarrow0\,,
\label{limit_massless}
\end{align}
it can be checked that the curvature invariants of the metric (\ref{metric_static}) all become constant everywhere. Thus the black hole disappears in this limit, and we are left with (a part of) AdS space-time, Minkowski space-time or dS space-time, in the case $b<1$, $b=1$ or $b>1$, respectively. For this reason, we can interpret $c$ as the mass parameter of the black hole, and (\ref{limit_massless}) corresponds to taking the mass to be zero.

This limit of vanishing $c$ implies that $y_1\rightarrow-\infty$. In terms of the rod structure, the would-be horizon at $(-1\,{\le}\,x\,{\le}\,{+1},y\,{=}\,{-\infty})$, represented by Rod 2, collapses to a single regular point in this limit. Rods 1 and 3 then merge to become the sole axis of the space-time. Note that an acceleration horizon is still present in the space-time in the case $b\ge 0$.

\subsubsection{Schwarzschild--dS/AdS black hole}
\label{sec_static_Sch}

The Schwarzschild--dS/AdS black hole can be recovered from metric (\ref{metric_static}) by taking the limits
\begin{align}
b\rightarrow\pm\infty\,,\qquad c\rightarrow0\,,
\end{align}
in an appropriate way. Specifically, we set
\begin{align}
b=-\frac{1}{\epsilon^2}\,,\qquad c=\frac{2m\epsilon}{\ell}\,,\qquad x=\cos\theta\,,\qquad y=-\frac{\ell}{\epsilon r}\,,\qquad t\rightarrow\frac{\epsilon}{\ell}\,t\,.
\label{limit_Sch-AdS}
\end{align}
If we take the limit $\epsilon\rightarrow 0$, we recover the familiar form of the Schwarzschild--AdS black hole:
\begin{align}
\dif s^2&=-\bigg(1-\frac{2m}{r}+\frac{r^2}{\ell^2}\bigg)\dif t^2+\frac{\dif r^2}{1-\frac{2m}{r}+\frac{r^2}{\ell^2}}+r^2(\dif\theta^2+\sin^2\theta\,\dif\phi^2)\,.
\end{align}
The Schwarzschild--dS black hole can similarly be recovered, if we take the limit $\epsilon\rightarrow i0$ instead. The quantities in (\ref{limit_Sch-AdS}) remain real, since recall that $\ell$ is imaginary in this case.

We note from the second equation in (\ref{limit_Sch-AdS}) that $c$ is related to the mass $m$ of the black hole in this scaling limit. This is consistent with our interpretation of $c$ as the mass parameter in Sec.~\ref{sec_static_massless}. On the other hand, $b$ can be interpreted as the acceleration parameter. The zero-acceleration limit, as we have just seen, corresponds to taking its magnitude to infinity.

\subsubsection{The C-metric}
\label{sec_static_C}

If we take the Ricci-flat limit
\begin{align}
b\rightarrow 1\,,
\label{limit_Ricci_flat}
\end{align}
while keeping $\ell^2(1-b)\equiv \varkappa^2$ constant, the metric (\ref{metric_static}) becomes 
\begin{align}
	\dif s^2&=\frac{\varkappa^2}{(x-y)^2}\bigg[G(y)\dif t^2-\frac{\dif y^2}{G(y)}+\frac{\dif x^2}{G(x)}+G(x)\dif \phi^2\bigg]\,,\cr
G(x)&=(1-x^2)(1+cx)\,.
\end{align}
This is equivalent to the factorised form of the C-metric first proposed in \cite{Hong:2003gx}. In this form of the metric, the black-hole horizon is at $y=-\frac{1}{c}$, while the acceleration horizon is at $y=-1$.

\subsubsection{Static black bottle}
\label{sec_static_bottle}

In the limit of maximum $c$:
\begin{align}
c\rightarrow1\,,
\end{align}
the third root of $P$ joins up with the one at $x=-1$ to form a double root at this point. Because of this, the axis represented by Rod 1 becomes infinitely far away from the other points of the space-time. Since the south pole of the black-hole horizon is attached to this axis, it is also pushed to infinity. The horizon then takes the shape of a bottle, with an infinitely long neck. Such ``black bottle'' solutions were studied in detail in \cite{Chen:2016rjt}. As is clear from Fig.~\ref{fig_param_space1}, this limit only exists if $b<1$, i.e., in AdS space-time.

\subsubsection{Space-time with no static region}
\label{sec_static_extremal1}

Recall that for fixed $c$, the maximum value that $b$ can take occurs when the two roots $y_1$ and $y_2$ coincide. This limit corresponds to the situation where the black-hole horizon coincides with the acceleration horizon, thus leaving a space-time with no static region.

The condition $y_1=y_2$ can be solved parametrically in terms of the double root itself. If we denote $y_*\equiv y_1$, then we have
\begin{align}
Q(y\,{=}\,y_*)=0\,,\qquad Q'(y\,{=}\,y_*)=0\,,
\end{align}
which can be solved to obtain
\begin{align}
b=-\frac{y_*^2(1+y_*^2)}{1-3y_*^2}\,,\qquad c=\frac{2y_*}{1-3y_*^2}\,,
\label{extremal_horizons_static_param}
\end{align}
where
\begin{align}
-\infty<y_*<-1\,.
\end{align}
This gives the curve (a) in Fig.~\ref{fig_param_space1}. Since this curve satisfies $b>1$, it lies entirely in the dS region of the parameter space.

\subsubsection{AdS black hole with extremal acceleration horizon}
\label{sec_static_extremal2}

It is also possible for the two roots $y_2$ and $y_3$ to coincide, which corresponds to the acceleration horizon becoming extremal. The condition $y_2=y_3$ can be solved parametrically in terms of the double root itself. If we denote $y_*\equiv y_3$, we in fact get the same solution (\ref{extremal_horizons_static_param}), but with a different parameter range:
\begin{align}
0<y_*<\frac{1}{3}\,.
\end{align}
This gives the dashed curve (b) in Fig.~\ref{fig_param_space1}. Note that it lies entirely in the AdS region of the parameter space. It separates the AdS black holes with acceleration horizons from those without acceleration horizons.

\section{Rotating accelerating black hole}
\label{sec_Crotating}

The metric describing a rotating and accelerating spherical black hole can be written as
\begin{align}
\dif s^2&=\frac{\ell^2(1-b)}{(x-y)^2}\bigg[
\frac{Q(y)}{1+ax^2y^2}(\dif{t}-\sqrt{a}x^2\dif{\phi})^2
-\frac{1+ax^2y^2}{Q(y)}\,\dif{y}^2\nonumber\\
&\hspace{0.75in}+\frac{1+ax^2y^2}{P(x)}\,\dif{x}^2
+\frac{P(x)}{1+ax^2y^2}(\dif{\phi}+\sqrt{a}y^2\dif{t})^2\bigg]\,,\nonumber\\
P(x)&=1+cx-(1-ab)x^2-cx^3-abx^4,\nonumber\\
Q(y)&=b+cy-(1-ab)y^2-cy^3-ay^4.
 \label{metric_crotating}
\end{align}
We note that the functions $P$ and $Q$ can also be written as
\begin{align}
P(x)=(1-x^2)(1+cx+abx^2)\,,\qquad  Q(y)=P(y)-(1-b)(1+ay^4)\,.
\label{PQ_alt}
\end{align}
This solution has four parameters: $\ell$, $a$, $b$ and $c$. The static solution (\ref{metric_static}) is recovered by setting $a=0$, so $a$ can be interpreted as a rotation parameter. The other three parameters $\ell$, $b$ and $c$ have the same interpretations as in the static case.

The metric (\ref{metric_crotating}) was obtained from the Pleba\'nski--Demia\'nski solution \cite{Plebanski:1976gy}, by imposing that (a) $P$ admits at least two real roots corresponding to the two axes of the space-time; and (b) the two axes have the same generator up to an overall normalisation. These two conditions ensure that the solution describes a spherical black hole free of NUT charge. In particular, (b) implies that the two roots of $P$ must have opposite sign to each other. By an appropriate rescaling of the coordinates, they can be fixed at $x=\pm 1$, and the metric (\ref{metric_crotating}) is obtained. This derivation is similar to that of the black bottle solution from the Pleba\'nski--Demia\'nski solution in \cite{Chen:2016rjt}, and interested readers may refer to Appendix A of that paper for more details.

\subsection{Coordinate and parameter ranges}

As in the static case, we take the range of $x$ to be (\ref{range_x}). Since $P(x\,{=}\,0)=1>0$, $P$ should remain positive within this range. This requires the larger factor of $P$ in (\ref{PQ_alt}) to be positive within this range. In particular, its non-negativity at the two endpoints $x=\pm1$ implies that
\begin{align}
-(1+ab)\leq c\leq1+ab\,.
\label{ex_axis}
\end{align}

We next require that the metric (\ref{metric_crotating}) is real-valued and has the correct Lorentzian signature. To ensure that it is real-valued, note that the parameter $a$ has to be non-negative:
\begin{align}
a\geq0\,.
\label{positive_a}
\end{align}
The requirement of Lorentzian signature means that $Q<0$ and $\ell^2(1-b)>0$, just as in the static case. In particular, the latter condition implies that $b<1$, $b=1$ and $b>1$, for the AdS, Ricci-flat and dS cases, respectively.

We now turn to the range of $y$. As in the static case, conformal infinity of the metric (\ref{metric_crotating}) lies at $x=y$. The reflection symmetry of the metric (c.f.\ Footnote \ref{reflection}) allows us to consider only the region with $y<x$. It can also be checked that (\ref{metric_crotating}) admits curvature singularities at $(x\,{=}\,0,y\,{=}\,{\pm\infty})$ and $(x\,{=}\,{\pm\infty},y\,{=}\,0)$. To avoid the curvature singularity at $(x\,{=}\,0,y\,{=}\,{-\infty})$, we demand that $Q$ admits at least one real root $y_1$ satisfying $y_1<-1$. This root can be interpreted as the event horizon of the black hole, which encloses the curvature singularity. By restricting $y$ to the range $y>y_1$, we will be focussing on the stationary region outside the black hole.

Since $Q<0$ in this region, it should satisfy the conditions (\ref{condition_existence_BH}). But now that $Q$ is a quartic function with a negative leading coefficient $-a$, it follows that $Q$ admits another real root $y_0$ satisfying
\begin{align}
-\infty<y_0\leq y_1<-1\,.
\label{Q_root_structure1}
\end{align}
The appearance of this root is not unexpected, since adding rotation to a black hole causes an inner horizon to appear. The root $y_0$ can be identified as the inner horizon of the black hole.

\subsubsection{Domain structure}

In this subsection, we will analyse the root structure of $Q$, and hence determine the possible domains for the metric (\ref{metric_crotating}). Apart from the appearance of the new root $y_0$, the qualitative behaviour of $Q$ is actually similar to that in the static case. We again consider the different cases given by the sign of the cosmological constant. 

In the AdS case for which $b<1$, $Q$ may admit either two or four real roots. If $Q$ admits only two real roots, the root structure is simply described by (\ref{Q_root_structure1}), and the domain of interest is (\ref{coor_range1}). If $Q$ admits four real roots, the two other roots lie in between $-1$ and $+1$. This follows from the fact that
\begin{align}
Q(y\,{=}\,{\pm1})&=-(1-b)(1+a)<0\,,\nonumber\\
Q'(y\,{=}\,{-1})&=2(1-c+ab)+4a(1-b)>0\,,\nonumber\\
Q'(y\,{=}\,{+1})&=-2(1+c+ab)-4a(1-b)<0\,,
\label{convex}
\end{align}
where the conditions (\ref{ex_axis}) and (\ref{positive_a}) have been used. If we denote these two roots as $y_{2,3}$, we have
\begin{align}
-\infty<y_0\le y_1<-1<y_2\le y_3<+1\,.
\label{coord_range_AdS}
\end{align}
The domain of interest is then (\ref{coor_range2}).

In the dS case for which $b>1$, it follows that
\begin{align}
Q(y\,{=}\,{\pm1})=(b-1)(1+a)>0\,.
\end{align}
Hence, $Q$ admits four real roots, satisfying
\begin{align}
-\infty<y_0\le y_1<y_2<-1<+1<y_3\,.
\end{align}
The domain of interest is then (\ref{coor_range4}). In the Ricci-flat case for which $b\rightarrow 1$, we see that $y_{2,3}\rightarrow\mp1$. So the four real roots of $Q$ satisfy
\begin{align}
-\infty<y_0\le y_1<y_2=-1<+1=y_3\,.
\end{align}
The domain of interest is then (\ref{coor_range3}).

Since the above four possible domains of interest are exactly the same as in the static case, they can be visualised as in Fig.~\ref{fig_domains}. The only difference is the appearance of the new root $y_0$, which is not shown in the plots.

\subsubsection{Parameter space}
\label{sec_param_space}

\begin{figure}
	\begin{center}
		\includegraphics[width=4.9in]{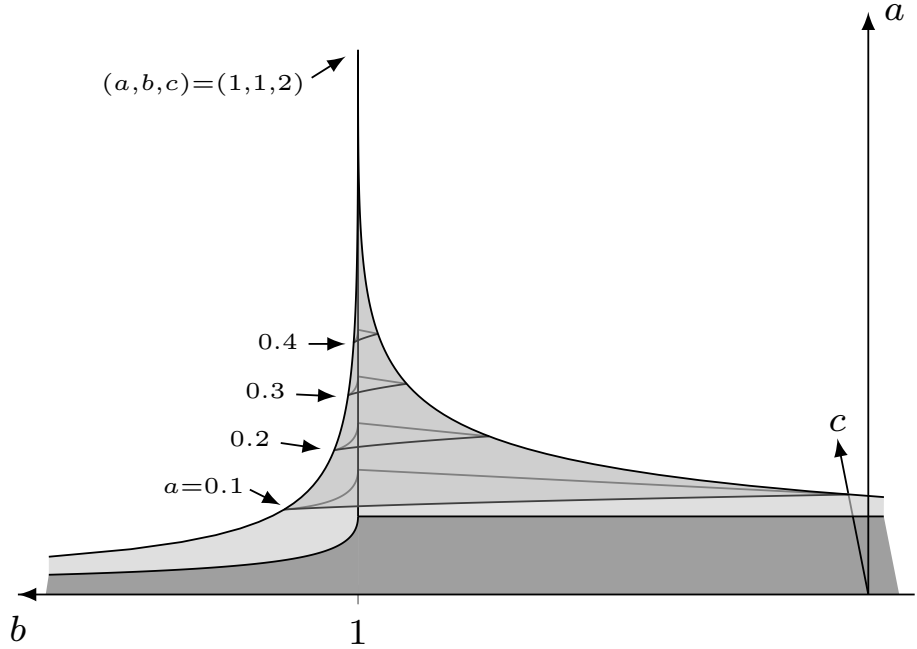}
	\end{center}
	\caption{The parameter space of the metric (\ref{metric_crotating}). It  has the general shape of a pyramid, whose height is parameterised by $a$. For clarity, only the part of the pyramid with $0.1\leq a<1$ is shown (in medium gray). The part with $0<a<0.1$ has been cut out to reveal the two far sides of the pyramid (in light gray) and its base (in dark gray). The base of the pyramid $a=0$ corresponds precisely to the parameter space in Fig.~\ref{fig_param_space1}.}
	\label{fig_param_space2}
\end{figure}

We now describe the $a$-$b$-$c$ parameter space of the metric (\ref{metric_crotating}). It turns out that the allowed range of parameters is more restrictive than (\ref{ex_axis}) and (\ref{positive_a}), since we have not fully exploited the restrictions on the parameters imposed by the condition (\ref{Q_root_structure1}). The full parameter space is determined only after all three conditions are solved for, and is shown in Fig.~\ref{fig_param_space2}.

\begin{figure}
	\begin{center}
		\includegraphics[width=6.4in]{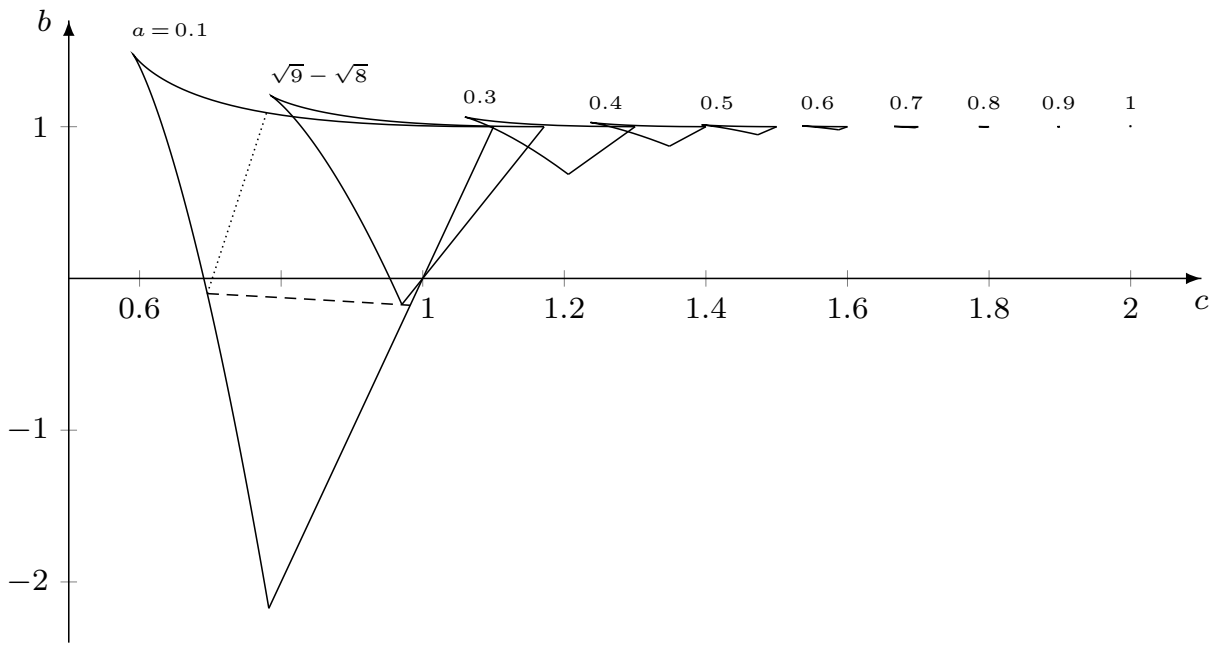}
	\end{center}
	\caption{Constant-$a$ slices of the parameter space of the metric (\ref{metric_crotating}) for various values of $a$ as indicated. Each slice has the general shape of a triangle. As $a$ is increased, the triangles become smaller and shift in the $c$-direction. They shrink to a point when $a=1$, corresponding to the apex of the pyramid in Fig~\ref{fig_param_space2}.}
	\label{fig_param_space3}
\end{figure}

As can be seen, the parameter space has the general shape of a pyramid, whose height is parameterised by $a$. Constant-$a$ slices of the pyramid are then parameterised by $b$ and $c$. The base of the pyramid, given by $a=0$, is nothing but the parameter space of the static accelerating black hole shown in Fig.~\ref{fig_param_space1}. Thus the base is unbounded in the $b$-direction, although it is bounded in the $c$-direction. For non-zero $a$ however, the slices of the pyramid are bounded in both the $b$- and $c$-directions. A series of constant-$a$ slices of the pyramid are shown in Fig.~\ref{fig_param_space3}.

Each of the constant-$a$ slices in Fig.~\ref{fig_param_space3} has the general shape of a triangle. It is clear that the area of the triangles decreases as $a$ is increased, and it vanishes when $a=1$. This defines the apex of the pyramid. It is also clear that the triangles shift in the $c$-direction as $a$ is increased, with the apex of the pyramid located at $(a,b,c)=(1,1,2)$. Thus the pyramid is slanted in the $c$-direction, although this is not apparent in Fig.~\ref{fig_param_space2} because of the perspective chosen. The maximum allowed value of both $a$ and $c$ are reached at the apex. This implies the global boundedness of the parameter $a$:
\begin{align}
0\le a<1\,,
\end{align}
and of the parameter $c$:
\begin{align}
0\le c<2\,.
\label{range_c_rotating}
\end{align}

Let us now study the constant-$a$ slices with non-zero $a$ in more detail. As can be seen from Fig.~\ref{fig_param_space3}, each of these slices is bounded by three curves, which can be unambiguously referred to as the top, bottom-left and bottom-right curves. They are the generalisations of the three boundary curves in Fig.~\ref{fig_param_space1} to non-zero $a$. We shall describe each one in turn.

The lower-right boundary curve is a straight line satisfying the equation $c=1+ab$. For this value of $c$, $P$ has a double root at $x=-1$. This has the effect of sending the south pole of the black-hole horizon to infinity. Thus, points on this line describe rotating black bottles. Note that this curve lies entirely in the AdS region of the parameter space. This special case will be discussed in Sec.~\ref{sec_bottle_rotating}.

The top boundary curve corresponds to the situation in which $y_1=y_2$. Thus, the Lorentzian region of interest vanishes along this curve, and no stationary space-time exists beyond it. Note that this curve lies entirely in the dS region of the parameter space. This special case will be discussed in Sec.~\ref{sec_rotating_extremal2}.

Lastly, the lower-left boundary curve corresponds to the situation in which $y_0=y_1$. Thus, points on this curve describe rotating black holes whose inner and outer horizons coincide, i.e., extremal rotating black holes. Since this curve passes through $b=1$, it lies in both the dS and AdS regions of the parameter space. This special case will be discussed in Sec.~\ref{sec_rotating_extremal1}.

For values of $a$ lying in the range
\begin{align}
0<a<\sqrt{9}-\sqrt{8}\,,
\label{range_a1}
\end{align}
the constant-$a$ slices contain two additional physically significant curves within them. They are depicted as the dashed and dotted curves in the $a=0.1$ slice in Fig.~\ref{fig_param_space3}. We now describe their significance.

The dashed curve corresponds to the situation in which $y_2=y_3$.  Thus, points on this curve describe rotating black holes whose acceleration horizons have become extremal. It is the generalisation of the dashed curve in Fig.~\ref{fig_param_space1} to non-zero $a$, and lies entirely in the AdS region of the parameter space. It divides the parameter slice into two: below it are solutions describing black holes without acceleration horizons; above it are solutions describing black holes with acceleration horizons. This special case will be discussed in Sec.~\ref{sec_rotating_extremal3}.

For solutions with both a black-hole and an acceleration horizon, it is natural to ask if thermal equilibrium can be achieved between them. This is precisely achieved on the dotted curve, which is actually a straight line. Since this line passes through $b=1$, it lies in both the dS and AdS regions of the parameter space. Solutions to the left of it have a black-hole horizon whose temperature is lower than that of the acceleration horizon, while the opposite is true for solutions to the right of it. This special case will be discussed in Sec.~\ref{sec_thermo_equi}.

When $a=0$, the dotted curve actually lies within the $c=0$ boundary of Fig.~\ref{fig_param_space1}. As $a$ is increased, it shifts in the $c$-direction. At the same time, the dashed curve also shifts slightly, and is shortened in length. At the critical value $a=\sqrt{9}-\sqrt{8}$, the dotted curve coincides with the lower-right boundary curve $c=1+ab$, while the dashed curve shrinks to a point which coincides with the bottom corner of the parameter slice. When $a$ is increased beyond this critical value, the dashed and the dotted curves vanish from the parameter slice. This means that there is always an acceleration horizon present in the space-time, and that its temperature is always higher than that of the black-hole horizon.

We remark that the Kerr-dS and Kerr-AdS black holes can be recovered in the scaling limit $b\rightarrow\pm\infty$, $c\rightarrow0$ and $a\rightarrow0$. In Fig.~\ref{fig_param_space2}, this corresponds to approaching the two infinite corners of the pyramid, where the $c=0$ line meets the other boundary curves. These two special cases will be discussed in Sec.~\ref{sec_Kerr_AdS}.

Finally, we note that for the above parameter space, the following inequality holds:
\begin{align}
-1<ab<1\,.
\label{range_ab}
\end{align}
Although it is not obvious, this result can be rigorously proved. It will be needed when we show that the space-time is free of CTCs in Sec.~\ref{sec_CTCs}.

\subsection{Geometrical and physical properties}

\subsubsection{Rod structure}
\label{section_rod_structure}

We now consider the rod structure of the metric (\ref{metric_crotating}). When an acceleration horizon is present, the rod structure is as follows:
\begin{enumerate}
	\item Rod 1: a semi-infinite space-like rod located at $(x\,{=}\,{-1},y_1\,{\leq}\,y\,{<}\,{-1})$, with direction
	\begin{align}
	k_1=\frac{1}{\kappa_{\text{E}1}}(\sqrt{a},1)\,,\qquad \kappa_{\text{E}1}=1-c+ab\,;
	\end{align}
	
	\item Rod 2: a finite time-like rod located at $(-1\,{\leq}\,x\,{\leq}\,{+1},y\,{=}\,y_1)$, with direction
	\begin{align}
	k_2=\frac{1}{\kappa_2}(1,-\sqrt{a}y_1^2)\,,\qquad \kappa_2=-\frac{1}{2}\frac{\dif Q}{\dif y}\bigg|_{y=y_1}\,;
	\end{align}
	
	\item Rod 3: a finite space-like rod located at $(x\,{=}\,{+1},y_1\,{\leq}\,y\,{\leq}\,y_2)$, with direction
	\begin{align}
	k_3=\frac{1}{\kappa_{\text{E}3}}(\sqrt{a},1)\,,\qquad \kappa_{\text{E}3}=1+c+ab\,;
	\end{align}
	
	\item Rod 4: a time-like rod located at $(\max(y_2,-1)\,{\leq}\,x\,{\leq}\,{+1},y\,{=}\,y_2)$, with direction
	\begin{align}
	k_4=\frac{1}{\kappa_4}(1,-\sqrt{a}y_2^2)\,,\qquad \kappa_4=\frac{1}{2}\frac{\dif Q}{\dif y}\bigg|_{y=y_2}\,.
	\end{align}
\end{enumerate}
It can be checked that the surface gravities in the above rod directions are all non-negative. When the acceleration horizon is absent, Rod 4 becomes irrelevant in the rod structure, as in the static case.

From the above rod structure, we note that Rods 1 and 3 have parallel directions. Thus, we can identify either $k_1$ or $k_3$ as the generator of the axial symmetry of the solution; either way, the axes described by Rods 1 and 3 are consistently defined, and NUT charge is absent from the space-time \cite{Chen:2010zu}. This confirms that the metric (\ref{metric_crotating}) is the natural rotating generalisation of the static metric (\ref{metric_static}).

\subsubsection{Temporal and azimuthal coordinates}

When $a$ is non-zero, note that the directions of Rods 1 and 3 are no longer purely along $\frac{\partial}{\partial\phi}$. This can be fixed by defining a new temporal coordinate $\tau$ by
\begin{align}
	\tau=t-\sqrt{a}\phi\,.
	\label{linear_transformation}
\end{align}
The metric (\ref{metric_crotating}) then becomes
\begin{align}
	\dif s^2&=\frac{\ell^2(1-b)}{(x-y)^2}\bigg\{
\frac{Q(y)}{1+a x^2y^2}\big[\dif \tau+\sqrt{a}(1-x^2)\dif \phi\big]^2-\frac{1+a x^2y^2}{Q(y)}\,\dif y^2
\nonumber\\
	&\hspace{0.788in}
+\frac{1+a x^2y^2}{P(x)}\,\dif x^2+\frac{P(x)}{1+a x^2y^2}\big[(1+ay^2)\dif \phi+\sqrt{a}y^2\dif\tau\big]^2
\bigg\}\,.
	\label{metric_Crotating_new_coordinates}
\end{align}
One can again calculate the rod structure in these coordinates. It turns out to be identical to the above rod structure, except that the rod directions are given in terms of the new basis $\big\{\frac{\partial}{\partial \tau},\frac{\partial}{\partial\phi}\big\}$ by
\begin{subequations}
\label{rod_structure_new_coordinates}
\begin{align}
k_1&=\frac{1}{\varkappa_{\text{E}1}}(0,1)\,,& &\hskip-2cm\varkappa_{\text{E}1}=1-c+ab\,;\\
\label{rod_structure_new_coordinates_2}
k_2&=\frac{1}{\varkappa_2}\left(1,-\frac{\sqrt{a}y_1^2}{1+ay_1^2}\right),& &\hskip-2cm\varkappa_2=-\frac{1}{2(1+ay_1^2)}\frac{\dif Q}{\dif y}\bigg|_{y=y_1}\,;\\
k_3&=\frac{1}{\varkappa_{\text{E}3}}(0,1)\,,& &\hskip-2cm\varkappa_{\text{E}3}=1+c+ab\,;\\
\label{rod_structure_new_coordinates_4}
k_4&=\frac{1}{\varkappa_4}\left(1,-\frac{\sqrt{a}y_2^2}{1+ay_2^2}\right),& &\hskip-2cm\varkappa_4=\frac{1}{2(1+ay_2^2)}\frac{\dif Q}{\dif y}\bigg|_{y=y_2}\,.
\end{align}
\end{subequations}
As desired, the directions of Rods 1 and 3 are now purely along $\frac{\partial}{\partial\phi}$. The axial symmetry of the space-time is thus generated by $\frac{\partial}{\partial\phi}$, so $\phi$ is the azimuthal coordinate of the space-time.

To avoid a conical singularity along Rod 1, the coordinate identification
\begin{align}
(\tau,\phi)\rightarrow (\tau,\phi+\Delta \phi_1)\,,\qquad \Delta\phi_1\equiv \frac{2\pi}{1-c+ab}\,,
\label{identification_rotating1}
\end{align}
has to be made. On the other hand, to avoid a conical singularity along Rod 3, the coordinate identification
\begin{align}
	(\tau,\phi)\rightarrow (\tau,\phi+\Delta\phi_2)\,,\qquad \Delta\phi_2\equiv\frac{2\pi}{1+c+ab}\,,
	\label{identification_rotating2}
\end{align}
has to be made. From the ranges of parameters we have identified, in particular from $0\leq c\leq1+ab$, we see that
\begin{align}
\Delta\phi_1\geq\Delta\phi_2\,,
\end{align}
with equality holding only if $c=0$. Thus, as in the static case, the identifications (\ref{identification_rotating1}) and (\ref{identification_rotating2}) in general cannot be made simultaneously. If we choose to avoid a conical singularity along $x=+1$, then there will necessarily be one along $x=-1$, and {\it vice versa\/}.

\subsubsection{Horizon geometries}

To study the black-hole horizon geometry, it is convenient to reparameterise the solution in terms of $a$, $c$ and $y_1$. This amounts to writing $b$ as
\begin{align}
	b=y_1^2+\frac{cy_1(y_1^2-1)}{1+ay_1^2}\,.
	\label{b_in_y1}
\end{align}
For a constant time slice, the black-hole horizon has the induced metric
\begin{align}
	\dif s^2_\text{BH}&=\frac{\ell^2(1-y_1^2)(1+cy_1+ay_1^2)}{(1+ay_1^2)(x-y_1)^2}\bigg[\frac{1+a x^2y_1^2}{P(x)}\,\dif x^2+\frac{P(x)}{1+a x^2y_1^2}(1+ay_1^2)^2\dif \phi^2\bigg]\,.
	\label{metric_horizon_Crotating}
\end{align}
It can be checked that this metric is positive semi-definite for our ranges of interest.

For definiteness, we will make the identification (\ref{identification_rotating2}) in the rest of this subsection. This makes the north pole of the horizon a regular point, but there will be a conical singularity at the south pole. The deficit angle at the south pole is
\begin{align}
\delta=\frac{4\pi c}{1+c+ab}\,.
\label{conical_deficit_rotating}
\end{align}
It can be checked that $0\le\delta\le2\pi$, with the lower bound being reached in the Kerr--dS/AdS limit (c.f.\ Sec.~\ref{sec_Kerr_AdS}), and the upper bound being reached in the black bottle limit (c.f.\ Sec.~\ref{sec_bottle_rotating}). With the identification (\ref{identification_rotating2}), the area of the horizon is
\begin{align}
A_{\text{BH}}&=-2 \ell^2(1+cy_1+ay_1^2)\Delta\phi\nonumber\\
&=-4\pi \ell^2\bigg(1+\frac{c(1-y_1)(1-ay_1)}{(1+ay_1^2)(1+cy_1+ay_1^2)}\bigg)^{-1}.
\end{align}

The scalar curvature of the metric (\ref{metric_horizon_Crotating}) is 
\begin{align}
R(x)=-\frac{2}{\ell^2}+\frac{2c(x-y_1)^3(1+ay_1^2)(1-3axy_1-3ax^2y_1^2+a^2x^3y_1^3)}{\ell^2(1-y_1^2)(1+cy_1+ay_1^2)(1+ax^2y_1^2)^3}\,.
\end{align}
Unlike the static case, it can be checked that the scalar curvature can become negative at the north pole of the horizon for certain parameter ranges. When this happens, it is no longer possible to embed this region of the horizon in a three-dimensional Euclidean space. In the cases when such an embedding is possible, they turn out to be qualitatively similar to those in Fig.~\ref{fig_horizon}.

We remark that the geometry of the acceleration horizon can be analysed along similar lines, and the results are qualitatively similar to the static case.

Now, recall that the second components in the brackets of $k_{2,4}$ in (\ref{rod_structure_new_coordinates}) represent the angular velocities of the black-hole and acceleration horizons, respectively. From these expressions and the inequality $y_1^2>y_2^2$, we see that
\begin{align}
	|\Omega_{\text{BH}}|>|\Omega_{\text{AH}}|\,.
\end{align}
So the two horizons always have different angular velocities: they can never be in dynamical equilibrium. The black-hole horizon is always rotating faster than the acceleration horizon. On the other hand, the temperatures of the black-hole and acceleration horizons do not have a definite order, and they can in fact be in thermal equilibrium (c.f.\ Sec.~\ref{sec_thermo_equi}).

\subsubsection{Absence of CTCs}
\label{sec_CTCs}

Recall that the metric (\ref{metric_Crotating_new_coordinates}) does not contain any NUT charge, which means that the axes at $x=\pm 1$ can be consistently defined with a common generator $\frac{\partial}{\partial\phi}$. The time coordinate $\tau$ in (\ref{metric_Crotating_new_coordinates}) remains non-compact and has the range $-\infty<\tau<\infty$. These properties, however, do not guarantee the absence of closed time-like curves (CTCs) in the space-time. The absence of CTCs requires not only that $\tau$ is non-compact, but also that the generator $\frac{\partial}{\partial\phi}$ does not vanish except on the axes themselves. This is in turn equivalent to the requirement that $g_{\phi\phi}$ in the metric (\ref{metric_Crotating_new_coordinates}) is non-negative in the domain of interest.

Direct computation yields
\begin{align}
	g_{\phi\phi}=\frac{\ell^2(1-b)(1-x^2)F}{(x-y)^2(1+a x^2y^2)}\,,
\end{align}
where $F$ is defined as
\begin{align}
	F\equiv (1+ab)(1+ay^2)(1+ax^2y^2)+c[x(1+a^2y^4)+ay(1+x^2y^2-(x-y)^2)]\,.
\end{align}
We now prove that $F$ is positive in the region
\begin{align}
-1<x<1\,,\qquad  y_1<y<x\,;
\end{align}
this is sufficient to rule out CTCs in all the four possible domains of interest in Fig.~\ref{fig_domains}. For simplicity, we assume $0<a<1$; it can be verified directly that the static case $a=0$ has no CTCs. Invoking the second inequality in (\ref{ex_axis}), we then have
\begin{align}
F&\geq c(1+ay^2)(1+ax^2y^2)+c[x(1+a^2y^4)+ay(1+x^2y^2-(x-y)^2)]\nonumber\\
&=c(1+x)H\,,
\end{align}
where $H$ is defined as
\begin{align}
H&\equiv (1-axy)(1-ay^3)+ay(1+y)(1+xy)\,.
\end{align}
Clearly, CTCs are absent if $H$ can be shown to be positive.

The proof that $H$ is positive is similar to that in \cite{Chen:2016rjt}. We begin by noting that $H$ is linear in $x$. Firstly, we consider the triangular range of coordinates satisfying
\begin{align}
-1\le y<x<+1\,.
\label{range1_CTC}
\end{align}
$H$ is positive in this range since
\begin{align}
H(x\,{=}\,y)=(1+ay)(1+ay^4)>0\,,\qquad H(x\,{=}\,{+1})=(1+ay^2)^2>0\,.
\end{align}
Now we consider the rectangular range of coordinates satisfying
\begin{align}
y_1<y<-1<x<+1\,.
\label{range2_CTC}
\end{align}
Since we have $H(x\,{=}\,{+1})>0$, it remains to show that
\begin{align}
K\equiv H(x\,{=}\,{-1})=1+2ay-2ay^3-a^2y^4>0\,,
\end{align}
for all $y_1<y<-1$. Note that $K$ is a quartic polynomial in $y$ with negative leading coefficient, and that
\begin{align}
K'(y\,{=}\,{-1})=4a(a-1)<0\,,\qquad K'(y\,{=}\,0)=2a>0\,.
\end{align}
So for $y_1<y<-1$, the minimum of $K$ lies either at $y=y_1$ or $y=-1$. In either case, $K$ is positive after some algebra:
\begin{align}
K(y\,{=}\,{-1})&=1-a^2>0\,,\nonumber\\
K(y\,{=}\,y_1)&=K(y\,{=}\,y_1)-aQ(y\,{=}\,y_1)\nonumber\\
&=(1-ab)(1+ay_1^2)+ay_1(2-c)(1-y_1^2)>0\,,
\end{align}
where the equation $Q(y_1)=0$ and the inequalities (\ref{range_c_rotating}) and (\ref{range_ab}) have been used. This concludes the proof that there are no CTCs in the domains of interest.

\subsection{Special cases}
\label{sec_special_cases}

The static accelerating black hole discussed in Sec.~\ref{sec_static} is obtained from the metric (\ref{metric_crotating}) by setting $a=0$. This subclass of solutions forms the bottom boundary of the parameter space shown in Fig.~\ref{fig_param_space2}. In this subsection, we will discuss several other special cases of (\ref{metric_crotating}). In the process, we will obtain a better understanding of the various boundaries and parts of the parameter space.

\subsubsection{Kerr--dS/AdS black hole}
\label{sec_Kerr_AdS}

The Kerr--dS/AdS black hole can be recovered from metric (\ref{metric_crotating}) by taking the limits
\begin{align}
b\rightarrow\pm\infty\,,\qquad c\rightarrow0\,,\qquad a\rightarrow0\,,
\end{align}
in an appropriate way. This corresponds to approaching the two infinite corners of the pyramid in Fig.~\ref{fig_param_space2}. Specifically, we set
\begin{align}
b=-\frac{1}{\epsilon^2}\,,\qquad c=\frac{2m\epsilon}{\ell}\,,\qquad a=\frac{\alpha^2\epsilon^2}{\ell^2}\,,\qquad x=\cos\theta\,,\qquad y=-\frac{\ell}{\epsilon r}\,,\qquad t\rightarrow\frac{\epsilon}{\ell}\,t\,.
\label{limit_Kerr-AdS}
\end{align}
If we take the limit $\epsilon\rightarrow 0$, we recover the familiar form of the Kerr--AdS black hole:
\begin{align}
\dif s^2&=-\frac{Q}{\rho^2}(\dif t-\alpha\cos^2\theta\,\dif\phi)^2
+\rho^2\bigg(\frac{\dif r^2}{Q}+\frac{\dif\theta^2}{P}\bigg)
+\frac{P}{\rho^2}\sin^2\theta\,(r^2\dif\phi+\alpha\dif t)^2,\nonumber\\
P&=1-\frac{\alpha^2}{\ell^2}\cos^2\theta\,,\qquad 
Q=(r^2+\alpha^2)\bigg(1+\frac{r^2}{\ell^2}\bigg)-2mr\,,\qquad
\rho^2=r^2+\alpha^2\cos^2\theta\,.
\label{metric_Kerr-AdS}
\end{align}
The Kerr--dS black hole can similarly be recovered, if we take the limit $\epsilon\rightarrow i0$ instead. The quantities in (\ref{limit_Kerr-AdS}) remain real, since $\ell$ is imaginary in this case.

We note from the third equation in (\ref{limit_Kerr-AdS}) that $a$ is related to the rotational parameter $\alpha$ of the black hole in this scaling limit. This is consistent with our interpretation of $a$ as the rotational parameter of (\ref{metric_crotating}).

\subsubsection{Rotating C-metric}

If we take the Ricci-flat limit (\ref{limit_Ricci_flat}) while keeping $\ell^2(1-b)\equiv \varkappa^2$ constant, the metric (\ref{metric_crotating}) becomes 
\begin{align}
\dif s^2&=\frac{\varkappa^2}{(x-y)^2}\bigg[
\frac {G(y)}{1+ax^2y^2}(\dif{t}-\sqrt{a}x^2\dif{\phi})^2-\frac{1+ax^2y^2}{G(y)}\,\dif{y}^2
\nonumber\\
	&\hspace{0.71in}
+\frac{1+ax^2y^2}{G(x)}\,\dif{x}^2+\frac{G(x)}{1+ax^2y^2}(\dif{\phi}+\sqrt{a}y^2\dif{t})^2
\bigg]\,,\nonumber\\
	G(x)&=(1-x^2)(1+cx+ax^2)\,.
	\label{metric_Ricci-flat_rotating}
\end{align}
This is equivalent to the form of the rotating C-metric given in \cite{Hong:2004dm}. The roots of the factor $1+cy+ay^2$ in the function $G(y)$ are the locations of the inner and outer black-hole horizons. For these horizons to exist, the parameters $a$ and $c$ have to satisfy $0\le a<1$ and $2\sqrt{a}\le c<1+a$.

\subsubsection{Rotating black bottle}
\label{sec_bottle_rotating}

The far-right side of the pyramid in Fig.~\ref{fig_param_space2} (equivalently, the lower-right edges of the constant-$a$ slices in Fig.~\ref{fig_param_space3}) corresponds to the limit
\begin{align}
c\rightarrow1+ab\,.
\label{bottle_limit}
\end{align}
In this case, $P$ admits a double root at $x=-1$. This is the rotating generalisation of the black bottle solution discussed in Sec.~\ref{sec_static_bottle}. It occurs only in AdS space-time, and is studied in detail in \cite{Chen:2016rjt}. In particular, we note that the parameter space identified in \cite{Chen:2016rjt} (c.f.\ Fig.~6 of that paper) is precisely the far-right side of the pyramid in Fig.~\ref{fig_param_space2}.

\subsubsection{Accelerating extremal Kerr--dS/AdS black hole}
\label{sec_rotating_extremal1}

For fixed $a$ and $b$, the minimum value that $c$ can take lies on the near side of the pyramid in Fig.~\ref{fig_param_space2} (equivalently, the lower-left edges of the constant-$a$ slices in Fig.~\ref{fig_param_space3}). This occurs when the inner and outer black-hole horizons coincide, corresponding to the black hole becoming extremal.  

In terms of the roots of $Q$, this occurs when the two roots $y_0$ and $y_1$ coincide. If we denote $y_*\equiv y_0=y_1$, then we have the solution
\begin{align}
\label{bc}
b=-\frac{y_*^2(1+y_*^2)+ay_*^4(3-y_*^2)}{1-3y_*^2-ay_*^2(1+y_*^2)}\,,\qquad c=\frac{2y_*(1+ay_*^2)^2}{1-3y_*^2-ay_*^2(1+y_*^2)}\,.
\end{align}
It remains to deduce the range of the parameter $y_*$. We have, from (\ref{Q_root_structure1}), that $-\infty<y_*<-1$. However, it turns out that the actual range of $y_*$ is more restrictive than this. For each fixed $0<a<1$, it is bounded from below by the limit (\ref{bottle_limit}). This lower bound is
\begin{align}
y_{\rm a}\equiv1-2\sqrt{\frac{a+1}{a}}\cos\bigg[\frac{1}{3}\arccos\bigg({-}\sqrt{\frac{a}{a+1}}\bigg)\bigg]\,.
\end{align}
On the other hand, it is bounded from above by the requirement that the other two roots of $Q$ are greater than $y_*$. This upper bound is
\begin{align}
y_{\rm b}\equiv-\bigg[\frac{2\sqrt{(a+5)(a+1)}}{a}\cos\bigg(\frac{1}{3}\arccos\frac{-(a^2+8a+11)}{(a+5)^{3/2}\sqrt{a+1}}\bigg)-\frac{2+a}{a}\bigg]^{\frac{1}{2}}.
\end{align}

This solution can thus be parameterised by $a$ and $y_*$ (in addition to $\ell$), with the ranges
\begin{align}
y_{\rm a}<y_*<y_{\rm b}\,,\qquad 0<a<1\,.
\end{align}
Note that $b(y_*\,{=}\,y_{\rm a})<1$ and $b(y_*\,{=}\,y_{\rm b})>1$. This means that the lower bound of $y_*$ occurs in the AdS case, while the upper bound occurs in the dS case. In general, the accelerating extremal black hole can exist in both the dS and AdS cases; it is the accelerating generalisation of the extremal Kerr--dS/AdS black hole.

\subsubsection{Space-time with no stationary region}
\label{sec_rotating_extremal2}

For fixed $a$ and $c$, the maximum value that $b$ can take lies on the far-left side of the pyramid in Fig.~\ref{fig_param_space2} (equivalently, the upper edges of the constant-$a$ slices in Fig.~\ref{fig_param_space3}). This occurs when the outer black-hole horizon coincides with the acceleration horizon, thus leaving a space-time with no stationary region. 

In terms of the roots of $Q$, this occurs when the two roots $y_1$ and $y_2$ coincide. If we denote $y_*\equiv y_1=y_2$, then we have the same solution (\ref{bc}) but with a different parameter range for $y_*$. The lower bound of $y_*$ is given by $y_{\rm b}$ obtained above. On the the other hand, the upper bound of $y_*$ is simply given by $-1$. It can be verified that for this range, $y_*$ lies between the other two roots of $Q$ as required.

This solution can thus be parameterised by $a$ and $y_*$, with the ranges
\begin{align}
y_{\rm b}<y_*<-1\,,\qquad 0<a<1\,.
\end{align}
Since $b>1$ within these ranges, this solution only occurs in dS space-time.

\subsubsection{Rotating AdS black hole with extremal acceleration horizon}
\label{sec_rotating_extremal3}

It is also possible for the two largest roots of $Q$ to coincide, which corresponds to the acceleration horizon becoming extremal. Note from the possible domains in Fig.~\ref{fig_domains} that this can only occur in the AdS case. If we denote $y_*\equiv y_2=y_3$, the solution is again given by (\ref{bc}) but with a different parameter range for $y_*$. We have, from (\ref{coord_range_AdS}), that $-1<y_*<1$. However, the actual range of $y_*$ is more restrictive than this. For fixed $a$, it is bounded from below by the requirement that the other two roots of $Q$ are real and less than $y_*$. This lower bound is
\begin{align}
y_{\rm c}\equiv\frac{-1-a+\sqrt{1+6a+a^2}}{2\sqrt{a}}\,.
\end{align}
On the other hand, it is bounded from above by the limit (\ref{bottle_limit}). This upper bound is
\begin{align}
y_{\rm d}\equiv1-2\sqrt{\frac{a+1}{a}}\cos\bigg[\frac{1}{3}\bigg(\arccos\sqrt{\frac{a}{a+1}}+\pi\bigg)\bigg].
\end{align}
In the context of the constant-$a$ slices of Fig.~\ref{fig_param_space3}, this solution describes a curve starting on lower-left edge and ending on the lower-right edge. An explicit example is the dashed curve depicted in the $a=0.1$ slice. It should be noted however that it is not always the case that $y_{\rm c}<y_{\rm d}$. While this is true for values of $a$ less than $\sqrt{9}-\sqrt{8}$, we actually have $y_{\rm d}<y_{\rm c}$ for values of $a$ greater than this. The solution does not exist in this case.

Hence, rotating black holes with extremal acceleration horizons are parameterised by $a$ and $y_*$ satisfying
\begin{align}
y_{\rm c}<y_*<y_{\rm d}\,,\qquad 0<a<\sqrt{9}-\sqrt{8}\,.
\end{align}
It can be checked that $b<0$, consistent with the fact that this configuration occurs only in the AdS case. These solutions form a surface which cuts through the parameter space of Fig.~\ref{fig_param_space2} (not shown there), and divides it into two: one part describing black holes without acceleration horizons, and the other describing black holes with acceleration horizons.

\subsubsection{Black hole in thermal equilibrium with its acceleration horizon}
\label{sec_thermo_equi}

In the case when an acceleration horizon is present, there is a possibility of the black-hole horizon having the same temperature as the acceleration horizon. The condition of thermal equilibrium in this case translates to the equality $\varkappa_2=\varkappa_4$. From (\ref{rod_structure_new_coordinates_2}) and (\ref{rod_structure_new_coordinates_4}), we have the condition
\begin{align}
\frac{1}{1+ay_1^2}\frac{\dif Q}{\dif y}\bigg|_{y=y_1}+\frac{1}{1+ay_2^2}\frac{\dif Q}{\dif y}\bigg|_{y=y_2}=0\,.
\label{condition_thermal_eq}
\end{align}
At $y=y_{1,2}$, $Q$ of course vanishes:
\begin{subequations}
\begin{align}
Q(y_1)=0\,,\label{Qy1}\\
\quad Q(y_2)=0\,.\label{Qy2}
\end{align}
\end{subequations}
We first solve (\ref{condition_thermal_eq}) and (\ref{Qy1}) for $b$ and $c$, and then substitute their solutions into (\ref{Qy2}). The following equation is obtained:
\begin{align}
(1-ay_1y_2)^2-a(y_1+y_2)^2=0\,,
\label{EQ_TE}
\end{align}
and a solution to this is
\begin{align}
y_2=\frac{1+\sqrt{a}y_1}{ay_1-\sqrt{a}}\,.
\end{align}
The other solution $y_2=\frac{1-\sqrt{a}y_1}{ay_1+\sqrt{a}}$ to (\ref{EQ_TE}) is discarded, since it does not correspond to the parameters within the region in Fig.~\ref{fig_param_space2}. The parameters $b$ and $c$ can now be expressed in terms of $y_1$ and $a$ as follows: 
\begin{align}
b&=\frac{y_1(1+\sqrt{a}y_1)[(1+a)y_1-\sqrt{a}(2-(1-a)y_1^2)]}{(1-\sqrt{a}y_1)(1-a+\sqrt{a}(1+a)y_1+2ay_1^2)}\,,\nonumber\\
c &=
\frac{2\sqrt{a}(1+ay_1^2)^2}{(1-\sqrt{a}y_1)(1-a+\sqrt{a}(1+a)y_1+2ay_1^2)}\,.
\label{surface_TE}
\end{align}
It can be checked that, for fixed $a$, there is in fact a {\it linear\/} relationship between $b$ and $c$:
\begin{align}
c=\frac{2\sqrt{a}(1+ab)}{1-a}\,.
\end{align}

It remains to deduce the parameter ranges of $a$ and $y_1$ in (\ref{surface_TE}). For fixed $a$, $y_1$ is bounded from below and above by the requirement that $y_0<y_1$ and $y_1<y_2$, respectively. These minimum and maximum values are given by
\begin{align}
y_{\rm e}\equiv-\frac{1+a+\sqrt{1+6a+a^2}}{2\sqrt{a}}\,,\qquad y_{\rm f}\equiv\frac{1-\sqrt{2}}{\sqrt{a}}\,.
\end{align}
In the context of the constant-$a$ slices of Fig.~\ref{fig_param_space3}, this solution describes a straight line starting on lower-left edge and ending on the upper edge. An explicit example is the dotted line depicted in the $a=0.1$ slice. It turns out that when $a=\sqrt{9}-\sqrt{8}$, this line coincides with the lower-left edge given by $c=1+ab$. If $a$ is increased beyond this value, the line lies outside the parameter space.

Hence, black holes in thermal equilibrium with their acceleration horizons are parameterised by $a$ and $y_1$ satisfying
\begin{align}
y_{\rm e}<y_1<y_{\rm f}\,,\qquad 0<a<\sqrt{9}-\sqrt{8}\,.
\end{align}
These solutions form a surface which cuts through the parameter space of Fig.~\ref{fig_param_space2} (not shown there), and divides it into two: one part describing black holes that are colder than their acceleration horizons, and the other describing black holes that are hotter than their acceleration horizons. Since this surface crosses $b=1$, thermal equilibrium can occur in both the dS and AdS cases. We remark that when $y_1=y_{\rm e}$, the solution describes an extremal black hole in equilibrium with an extremal acceleration horizon. This zero-temperature configuration is parameterised by
\begin{align}
b=-a\,,\qquad c=2\sqrt{a}(1+a)\,,
\end{align}
and can only occur in the AdS case.

\section{Discussion}

In this paper, we have presented a new form of the rotating C-metric with cosmological constant (\ref{metric_crotating}). This solution describes the entire class of spherical black holes undergoing rotation and acceleration in dS or AdS space-time, including the class of ``slowly accelerating'' black holes in AdS space-time. When the cosmological constant vanishes, the form of the rotating C-metric in \cite{Hong:2004dm} is recovered. Unlike previous forms of this solution, the form presented here is simpler and allows for a complete characterisation of the parameter space. 

There are a few possible avenues for future work. One immediate generalisation of the solution (\ref{metric_crotating}) is to add an electric charge $e$ and a magnetic charge $g$ to it. The metric is given by
\begin{align}
\dif s^2&=\frac{\ell^2(1-b)}{(x-y)^2}\bigg[\frac{Q(y)}{1+a x^2y^2}(\dif t-\sqrt{a}x^2\dif \phi)^2-\frac{1+a x^2y^2}{Q(y)}\,\dif y^2\cr
&\hspace{0.75in}+\frac{1+a x^2y^2}{P(x)}\,\dif x^2+\frac{P(x)}{1+a x^2y^2}(\dif \phi+\sqrt{a}y^2\dif t)^2\bigg]\,,\nonumber\\
P(x)&=1-q^2+cx-(1-ab-2q^2)x^2-cx^3-(ab+q^2)x^4,\nonumber\\
Q(y)&=b-q^2+cy-(1-ab-2q^2)y^2-cy^3-(a+q^2)y^4,
\label{metric_charged}
\end{align}
and the corresponding gauge potential is given by
\begin{align}
{\cal A}=\frac{\sqrt{\ell^2(1-b)(1+a)}}{1+ax^2y^2}\left[ey(\dif t-\sqrt{a}x^2\dif\phi)-gx(\dif \phi+\sqrt{a}y^2\dif t)\right].
\label{Maxwell_rotating}
\end{align}
Here, we have defined $q\equiv \sqrt{e^2+g^2}$. Note that $P$ and $Q$ can also be written as
\begin{align}
P(x)=(1-x^2)[1-q^2+cx+(ab+q^2)x^2]\,,\qquad Q(y)=P(y)-(1-b)(1+ay^4)\,.
\end{align}
This solution describes a charged, rotating and accelerating black hole with cosmological constant. It can be analysed using methods similar to those employed in this paper.

Our main focus in this paper has been on black holes with spherical horizons. However, in the AdS case and for appropriately chosen coordinate and parameter ranges, the solution we have identified (\ref{metric_crotating}) (or (\ref{metric_charged})) can alternatively be interpreted as describing black holes with asymptotically hyperbolic horizons. In the static limit, the entire class of such solutions was identified in \cite{Chen:2015zoa} using the form (\ref{newform2}), in which the functions $P$ and $Q$ are assumed to have one real root each. It would be interesting to similarly reparameterise (\ref{metric_crotating}) and study the rotating generalisation of (\ref{newform2}).

Now, the Pleba\'nski--Demia\'nski solution generalises the solution (\ref{metric_charged}) with the inclusion of NUT charge, and represents the most general solution of type D in Petrov's classification. Although the presence of NUT charge leads to the existence of CTCs for spherical black holes, its interpretation in the case of hyperbolic black holes is less clear and could potentially be interesting. Thus we believe there is still merit to study this solution, to find a suitable parameterisation of it and identify the different classes of space-times contained within it. We remark that an alternative form of the Pleba\'nski--Demia\'nski solution has been proposed and studied in \cite{Griffiths:2005qp,Griffiths:2009dfa}. Nevertheless, it might be worthwhile to find a form of this solution that directly generalises the one proposed in this paper.

\section*{Acknowledgement}

This work was partially supported by the Academic Research Fund (WBS No.: R-144-000-333-112) from the National University of Singapore.

\bigskip\bigskip

{\renewcommand{\Large}{\normalsize}
}

\addtocontents{toc}{\protect\setcounter{tocdepth}{0}}

\end{document}